\documentclass[fleqn,usenatbib]{mnras}
\usepackage[pagewise]{lineno}
\usepackage{newtxtext,newtxmath}
\usepackage[T1]{fontenc}
\usepackage{indentfirst}

\DeclareRobustCommand{\VAN}[3]{#2}
\let\VANthebibliography\thebibliography
\def\thebibliography{\DeclareRobustCommand{\VAN}[3]{##3}\VANthebibliography}

\usepackage{graphicx}	% Including figure files
\usepackage{amsmath}	% Advanced maths commands
\usepackage{subcaption}

\title[Dynamics around a body with a crater]{Dynamics around non-spherical symmetric bodies – III. The case of a
spherical body with a crater}

\author[P.~V.~S.~Soares et al.]{
P.~V.~S.~Soares,$^{1}$\thanks{E-mail: paulo.v.soares@unesp.br}
G.~Madeira,$^{2}$\thanks{E-mail: madeira@on.br}
S.~M.~Giuliatti~Winter,$^{1}$
T.~Ribeiro$^{1}$
and O.~C.~Winter$^{1}$
\\
$^{1}$ Grupo de Dinâmica Orbital e Planetologia (GDOP), São Paulo State University-UNESP, Guaratinguetá, SP, CEP~12516-410, Brazil\\
$^{2}$ Observat\'orio Nacional/MCTI, R. General Jos\'e Cristino 77, CEP~20921-400, RJ, Brazil
}

\date{Accepted 2026 October 2. Received 2026 September 30; in original form 2026 June 17}

\pubyear{\the\year{}}

\begin{document}
\label{firstpage}
\pagerange{\pageref{firstpage}--\pageref{lastpage}}
\maketitle

\begin{abstract}

Motivated by the peculiar features of the trans-Neptunian object (TNO) Máni and the discovery of ring systems around Centaurs and trans-Neptunian bodies, we investigate the stability of particles around a small irregular body hosting a deep equatorial crater. This study is the third contribution in a sequence devoted to the dynamics around non-spherically symmetric bodies. Using Máni as a reference model, whose crater depth exceeds 10\% of its radius, we map the system stability through complementary methods: Poincaré surfaces of section (PSS), survival maps, and the finite-time Lyapunov exponent (FTLE). In the nominal case, the $1\!:\!1$, $2\!:\!1$, $3\!:\!1$, and $4\!:\!1$ spin-orbit resonances (SORs) are consistently identified by all techniques. The prominent $3\!:\!1$ SOR exhibits a bifurcated structure at high eccentricities, matching structural transitions observed at lower Jacobi constant values in the PSSs. Increasing the rotation rate ($\lambda$) and crater mass ratio ($\mu$) enlarges resonance widths and promotes overlap, making the $2\!:\!1$ and $4\!:\!1$ SORs progressively dominant. Lower rotation rates allow stable trajectories to persist at higher eccentricities, whereas the $1\!:\!1$ SOR is highly sensitive to stronger perturbations and disappears in the most extreme cases. In contrast to mass-anomaly models, which can clear the region interior to the $2\!:\!1$ SOR, the crater configuration considered here preserves stable regions in this vicinity, suggesting alternative scenarios for the formation and maintenance of debris structures around irregular minor bodies.

\end{abstract}

%%%%%%%%%%%%%%%%%%%%%%%%%%%%%%%%%%%%%%%%%%%%%%%%%%

%%%%%%%%%%%%%%%%% BODY OF PAPER %%%%%%%%%%%%%%%%%%

\section{Introduction}

Over a decade ago, stellar occultation observations revealed that ring systems are not exclusive to giant planets, but also exist around small solar system bodies. The first detection occurred around the Centaur (10199)~Chariklo \citep{volk2025centaurs}. Subsequently, evidence of ring material was reported for another Centaur, (2060)~Chiron \citep{ortiz2015possible}, with the scope of these discoveries later reaching massive Trans-Neptunian Objects (TNOs) such as the dwarf planet Haumea \citep{ortiz2017size} and Quaoar \citep{morgado2023dense}.

These small bodies frequently exhibit irregular, non-spherical shapes, generating perturbations in the gravitational potential around them that favour the formation of stability regions. Although modelling this potential had already been established through classical methods such as mass concentrations \citep[mascons, ][]{muller1968mascons} and polyhedrons \citep{werner1994gravitational, werner1996exterior}, these discoveries prompted further analyses dedicated to orbital dynamics around non-spherical bodies. Such studies revealed that gravitational field asymmetries excite spin-orbit resonances (SORs), which are capable of promoting the radial confinement of particles. This mechanism is explored in \cite{sicardy2019ring}, where the authors model Haumea and Chariklo as homogeneous triaxial ellipsoids and additionally consider Chariklo as a sphere with an equatorial topographic anomaly. Their results demonstrate that torques exerted by these non-axisymmetric features effectively sweep a disk of material that extends through the corotation SOR region. Under these conditions, the non-spherical gravitational field prevents material from remaining interior to the $2\!:\!1$ SOR on short time scales, explaining the clearing of this region prior to the formation of the narrow rings observed today. Thus, the existence of rings in this mechanism requires the central body to rotate fast enough so that the $2\!:\!1$ SOR is located interior to the Roche limit. This creates a stable zone between the SOR and the Roche limit where material can be confined without being swept away by torques or accreting into moons. 

This requirement is satisfied by the short rotation periods of Chariklo ($\sim$7h, \citealt{fornasier2014centaur}) and Haumea ($\sim$3.9h, \citealt{rabinowitz2006photometric}). Consequently, second-order SORs, such as the $3\!:\!1$ and $7\!:\!5$, emerge as the preferred environments for the confinement of the Chariklo, Haumea, and Quaoar ring systems \citep{leiva2017size, ortiz2017size}. Numerical simulations and stability analyses \citep{sicardy2025rings, celletti2026dynamical, salo2026rings} reinforce these constraints, revealing that the region interior to the $2\!:\!1$ SOR excites high eccentricities, rendering the long-term survival of rings unfeasible in that zone. Regarding the local dynamics of these stable regions, however, \citet{giuliatti2023stability} showed that Chariklo's inner ring (C1R) does not lie within the $3\!:\!1$ SOR; instead, due to its low eccentricity, it is associated with a distinct family of stable, non-resonant trajectories known as first-type orbits.

The present work follows the investigations of \cite{Madeira2022} and \cite{Ribeiro2023}, extending their analysis of non-axisymmetric bodies to the specific case of a body with a crater. \cite{Madeira2022} explored the dynamics around a spherical body with an equatorial mass anomaly ($\mu$), demonstrating that the instability in the inner region is primarily driven by the rotation parameter ($\lambda$). While they identified a chaotic region extending beyond the corotation radius for Chariklo, their Poincaré sections (e.g., their Fig. 13) reveal that stable structures, such as the $2\!:\!1$ SOR, can still emerge depending on the specific combination of the rotation period and size of the anomaly. 

Complementing this, \cite{Ribeiro2023} analysed Jacobi ellipsoids, using $J_2$ and $C_{22}$ coefficients to map stability. They found that while high ellipticity and fast rotation (as in Haumea) favour collisions and ejections, bodies with slower rotation allow for the stabilisation of resonant structures, including the $4\!:\!2$ SOR. Notably, they demonstrated that, for high values of $C_{22}$, traditional analytical methods fail to accurately locate SORs, requiring a geometrical approach which they termed "geometry of the orbit". Our study adopts a similar numerical methodology to evaluate how a localised crater shapes the phase space. Unlike \cite{Ribeiro2023}, the absence of a global $C_{22}$ term in our model allows for a more direct correlation with the analytical SOR positions, a characteristics that was also demonstrated for the case of a localised mass anomaly by \citet{Madeira2022}.

Our analysis also examines how the overall stability of the system changes due to the increasing degree of perturbation, driven in this case directly by both the crater size and the rotation of the central body, with the latter increasing the frequency of close encounters with the crater, which excites orbital eccentricity and modifies particle survival times. By focusing this investigation on the local dynamical structure, we characterise the individual $1\!:\!1$, $2\!:\!1$, $3\!:\!1$, and $4\!:\!1$ SORs via the Poincaré Surfaces of Section (PSS) technique for the nominal case based on Máni's parameters. Following this, the evolution of the system is tracked as the crater dimensions and the rotation period of the central body vary, allowing us to observe how these structural domains behave under enhanced perturbations.

The motivation for this scenario lies in observational discoveries concerning the TNO Máni (\(2002~MS_4\)). Based on nine stellar-occultation events, \cite{Rommel_2023} revealed a surface topography characterised by three prominent features: an approximately 11-km-deep depression, an adjacent elevation \(25^{+4}_{-5}\) km high, and a large crater-like depression extending \(322\pm39\) km and reaching \(45.1\pm1.5\) km in depth (Table~\ref{tab:ocultacoes_MS4}). These features were identified along the projected limb of Máni; however, their surface latitudes remain unconstrained because the orientation of the body’s rotation axis is not known with sufficient accuracy. The largest depression reaches a depth corresponding to more than 10\% of the object's radius and represents the largest crater-like feature observed relative to the size of its host body. In the present work, we deliberately simplify this observed topography by representing Máni as an otherwise idealised body containing a single dominant crater located at the equator. The equatorial location is therefore a modelling assumption rather than a constraint derived from the observations. This approximation allows us to isolate the dynamical effects of the dominant crater and the body’s rotation in its immediate vicinity.

%The motivation for this scenario lies in observational discoveries about the TNO Máni ($2002~MS_4$).  Detailed by \cite{Rommel_2023} from a campaign of nine stellar occultations, this body revealed a complex topography dominated by a vast equatorial crater 322 km wide and 45 km deep \ref{tab:ocultacoes_MS4}. Representing more than 10\% of the object's radius, this crater is the largest ever observed in proportion to the size itself. Such a prominent feature imposes a significant symmetry break in the gravitational potential.

%Os procedimentos teóricos são baseados na integração das equações de movimento do problema de dois corpos com rotação do corpo central \citep{SCHEERES199667} e a análise de estabilidade a partir do tempo de sobrevivência e do indicador de caos, sendo o lyapunov o indicador utilizado nesse trabalho.

%O transnetuniano (307261)~2002~MS\textsubscript{4}, atualmente chamado de Máni encontra-se no Cinturão de Kuiper, uma região além da órbita de Netuno que se estende aproximadamente de 37 a 48 unidades astronômicas (UA) do Sol. Essa região é de grande interesse científico, pois abriga os objetos transnetunianos (TNOs), considerados remanescentes da formação do Sistema Solar e que preservam informações sobre o período de formação do sistema solar externo \cite{barucci2008solar}. \gus{Acho que isso pode ser melhor fundamentado para o contexto desse trabalho e incluído na introdução}

Section \ref{secao2} establishes the theoretical and methodological basis of this work. In it, we define the dynamic model adopted and the equations of motion in the rotating frame, in addition to detailing the numerical tools used for stability analyses: PSS, survival time maps, and the calculation of Lyapunov Characteristic Exponent (LCE). Section \ref{Secao3} applies these methodologies to a model based on the TNO (307261) Máni. While Máni exhibits complex topographical features, including two craters and a mountain, our study focuses on a simplified configuration featuring a single equatorial crater. Using physical parameters derived from recent observations as a reference, we map the regions of stability and identify the main resonant structures of the system through the aforementioned techniques. In Section \ref{Secao4}, we expand the study by exploring the effects of varying physical parameters on the system's stability and dynamics. Specifically, we evaluate how the dynamical structure and the extent of stable regions evolve as a function of the body's rotation period and the crater's mass ratio. 
Finally, Section \ref{Secao5} presents the general conclusions and discussion of the results, summarising the dynamic implications observed.
%Section~5 establishes a critical comparative analysis between the results obtained for the crater and the anomalous mass and triaxial ellipsoid scenarios investigated, respectively, by \cite{Madeira2022} and \cite{Ribeiro2023}. The discussion focuses on identifying the predominant spin-orbit resonances in each model and the structures of the stability regions \gus{Não tem nada na seção 5}. Finally, Section~6 presents the general conclusions and discussion of the results, summarising the dynamic implications observed and suggesting perspectives for future work. \gus{Ao citar as seções, use os labels dela. Assim garante que sempre estará certo.}

%\gus{Acho que você poderia citar sim mais resultados dos artigos meu e da Tais, porém não acho essencial. O que você precisa tomar cuidado é com essa coisa de não existir material até a 2:1, do jeito que você escreveu isso parece uma verdade absoluta, enquanto no meu artigo eu mostro casos em que a 2:1 existe (Figura 13 mostra uma SSP da 2:1 por exemplo). E se eu me recordo, a Tais tbm encontra casos em que a 4:2 é estável.}

\section{Methods} \label{secao2}

In this work, we investigate the dynamics around a spherical body hosting a crater, with the aim of characterising stable, unstable, and chaotic regimes around this class of objects. As a reference case, we construct the central body using the physical parameters of the TNO Máni, presented in Table~\ref{tab:ocultacoes_MS4}. We also evaluate the survival of test particles in response to variations in the size and rotation ratio of the crater. This section describes the physical model and the numerical methods employed. We also present the PSS and the LCE, which are used to explore the stability of the system.

\subsection{Equations of Motion} \label{secao2.1}

In our modelling, we assume a system composed of two bodies: the central body, spherical with constant mass distribution — which we will refer to as the spherical portion of the body — in which an equatorial crater is dug into its surface, and a particle. The dynamics is described in a rotating frame centered on the body's center of mass, with the $z$-axis aligned with the constant rotation axis. The central body rotates on this axis without wobbling at an angular velocity $\omega = 2\pi / T$, where $T$ is the rotation period of the central body in seconds. We define the $x$-axis as the line connecting the center of mass to the equatorial crater, while the $y$-axis completes the right-handed orthogonal system. 

To simplify the system, physical quantities are normalised by adopting the gravitational constant $G = 1$, the mass of the central body and physical radius $M = 1$ and $R_c = 1$, respectively. Here, $R_c$ also represents the distance between the geometric centre of the spherical portion of the body and the crater. The unit of time is normalised using the critical Keplerian frequency \citep[see][]{lages2017chaotic, Madeira2022}, given by:

\begin{equation}
    \omega_k = \sqrt{\frac{GM}{R_c^3}} = \sqrt{\frac{4\pi G \rho}{3}} = 1
    \label{eq:freq.keplerianacritica}
\end{equation}
where $\rho$ is the density of the central body. $\omega_k$, determined by Kepler's third law, represents the angular velocity that a particle would have when orbiting the central body in a trajectory immediately above its surface.

The physical system is characterised by two parameters. The first is the body's rotation rate, given by $\lambda = \omega / \omega_k$. The second is the crater-body mass ratio parameter, defined by $\mu = m/M$, where $m$ represents the mass of the material removed to form the crater, and $M = M_0 + m$ is the total mass of the system ($M_0$ being the mass of the original spherical portion). Since a crater corresponds to a localised deficit of mass, we strictly define $\mu < 0$ for the crater case, where this negative value serves as a mathematical input in our numerical simulation codes to represent the removed mass. Conversely, a positive value ($\mu > 0$) corresponds to a mass anomaly \citep{sicardy2019ring, Madeira2022}. For physical clarity and to avoid the continuous use of negative parameters throughout the text, we present our results using the positive parameter $\overline{\mu} = |\mu|$, which represents the absolute magnitude of the mass ratio. Therefore, a larger value of $\overline{\mu}$ directly implies a deeper and more massive crater layout.

\begin{figure}
    \centering
    \includegraphics[width=1.\linewidth]{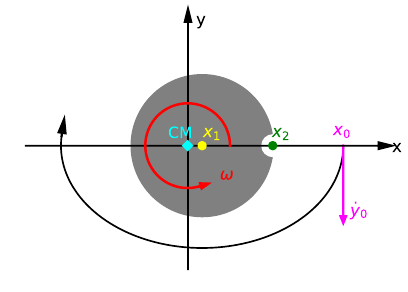}
    \caption{Scheme of the dynamical system. The central body is composed of a spherical portion (in gray) with its centre located at \(x_1\) and an equatorial crater (in green) with its centre at \(x_2\). The centre of mass (CM) of the complete body is shifted to the left with respect to the centre of the spherical portion and coincides with the origin of the reference frame. The body rotates with constant angular velocity \(\omega\). In magenta, we show the initial condition of a test particle at \(x_0\), with initial velocity \(\dot{y}_0\).}
    \label{fig:Esquema}
\end{figure}

Figure~\ref{fig:Esquema} shows the representation of the dynamic system considered in the numerical simulations. The centre of the spherical portion is indicated by the yellow dot, located at $x_1$, while the crater, indicated in green, is located at $x_2$. The centre of mass (CM) of the body, represented in turquoise blue, corresponds to the origin of the reference system and is displaced to the left of the centre of the spherical portion due to the removal of mass. The body rotates around its CM with angular velocity $\omega$, as indicated by the red arrow. With the adopted normalisation, the positions of the centres of mass of the components are $x_1 = -\mu$ and $x_2 = 1 - \mu$.

In this work, we restrict the analysis to the planar case, in which the motion of the test particle occurs exclusively in the equatorial plane of the central body. The position of the particle is described by $r = (x, y)$, starting from the initial conditions $(x_0, 0) \, R_c$ and velocity $(0, \dot{y}_0) \, R_c \omega_k$, as illustrated in Figure~\ref{fig:Esquema}. The expression for the gravitational potential in this system is given by:

\begin{equation}
U(\mathbf{r}) = -\frac{1}{\sqrt{(x-x_1)^2 + y^2}} - \frac{\mu}{\sqrt{(x-x_2)^2 + y^2}}
\label{Potencial}
\end{equation}
\noindent where the second term represents the contribution of the crater.

In comparison with the gravitational potential around a spherical body with a mass anomaly, presented in \cite{sicardy2019ring, Madeira2022}, it can be observed that Equation~\ref{Potencial} does not include the indirect term proportional to $\lambda^2$. This difference stems from the fact that the system is already formulated in a barycentric reference frame -- centred on the centre of mass of the central body and not on the centre of the spherical portion of the central body, as in \cite{sicardy2019ring, Madeira2022} -- making the introduction of the corrective term unnecessary.

The equations of motion in the rotational frame are given by  \citep{SCHEERES199667}:
\begin{equation}
\ddot{x} = 2\lambda \dot{y} + \lambda^2 x - F_x,
\label{eq:mov_x}
\end{equation}
\begin{equation}
\ddot{y} = -2\lambda \dot{x} + \lambda^2 y - F_y
\label{eq:mov_y}
\end{equation}
where $\dot{x}$ and $\dot{y}$ represent the components of the particle's velocity, and $\ddot{x}$ and $\ddot{y}$ represent the components of its acceleration. $F_x$ and $F_y$ stand for the partial derivatives of potential, $\partial U(\mathbf{r})/\partial x$ and $\partial U(\mathbf{r})/\partial y$, respectively.

The numerical integration of the equations of motion is performed using the \texttt{solve-IVP} function from the Python library SciPy \citep{bressert2012scipy} using the explicit Runge-Kutta method with adaptive step size, integrator RK45 \citep[see][]{dormand1980family}. As standard, we adopt relative and absolute tolerances equal to $10^{-8}$, ensuring calibrated accuracy in the calculation of trajectories.

\subsection{Spin-orbit resonances}

Spin-orbit resonance (SOR) occurs when a condition of commensurability is satisfied between the orbital frequency of a particle and the rotation frequency of the central body. The general condition for a SOR $m:(m-j)$ can be expressed as:
\begin{equation}
    m~n = (m - j)\omega_p + j\dot{\varpi}
    \label{spinorbita}
\end{equation}
where $m$ and $j$ are integers that define the commensurability ratio, and $j$ defines the SOR order. $n$ is the angular frequency of the particle and $\dot{\varpi}$ is the derivative of the particle’s longitude of pericentre. In this work, we consider only non-synchronous SORs -- sectorial resonances -- so we do not deal with the case of 1:1 SOR ($m=1$ and $j=0$), nor with cases where $j=m$, which correspond to apsidal resonances, i.e., resonances in which the particle's motion is commensurate with the precession rate of its line of apsides.

The temporal variation of the pericentre longitude is computed as $\dot{\varpi}=n-\kappa$, where the angular frequency $n$ and radial frequency $\kappa$ are given by \citep[see][]{Chandrasekhar,Madeira2022}:
\begin{equation}
    n^2 = \frac{1}{a^3}\left(1 + \mu + \frac{3\mu}{4}a^{-2}\right),
    \label{wpexpansion}
\end{equation}
and
\begin{equation}
    \kappa^2 = \frac{1}{a^3}\left(1 + \mu - \frac{3\mu}{4}a^{-2}\right)
    \label{kappaexpansion}
\end{equation}

\noindent where $a$ corresponds to the semi-major axis of the particle.

\subsection{Poincaré Surface of Section technique}

Used as a tool to analyse stable and chaotic regions, the PSS technique was originally designed for the analysis of the Restricted Three-Body Problem \citep{Henon1965a, Henon1965b, Henon1966a, Henon1966b, 1971Jefferys, Winter1994a, Winter1994b}, and was later modified for the study of the two-body problem with rotation \citep{SCHEERES199667}. In the context of the Two-Body Problem with asymmetry of the primary body, the technique was previously employed by \cite{Borderes2017, Winter2019, Madeira2022, Ribeiro2023,Madeira2025d} to different systems, such as systems with triaxial bodies or with mass anomaly, and is extended here to the case of a spherical body with a crater.

To obtain the PSS, we determine the state variables ($x, y, \dot{x}, \dot{y}$) by integrating equations \eqref{eq:mov_x} and \eqref{eq:mov_y}. For this, we assume a fixed value for the Jacobi constant $C_J$, a constant of motion expressed by \citep{SCHEERES199667}:
\begin{equation}   
    C_J = \lambda^2 (x^2 + y^2) + 2 U(\mathbf{r}) - \dot{x}^2 - \dot{y}^2 \label{CalculoCJ}
\end{equation}
 from which the initial velocity component $\dot{y}_0$ is obtained. For all trajectories investigated in this work, we adopt the initial condition $\dot{x}_0 = 0$, which implies that the particles begin their motion at one of the apsides. Specifically, due to the choice of the sign $\dot{y}_0<0$, the particles are integrated strictly from the apocentre. This standard configuration is uniformly applied across all resonant structures mapped in this study.

%However, for the characterisation of even-order SORs, such as the $3\!:\!1$ SOR -- the only second-order SOR analysed in this work --, we impose conditions with $\dot{x}_0 \neq 0$. This choice is justified by the fact that the structures associated with these SORs in the PSS exhibit symmetry with respect to the axis $\dot{x} = 0$; thus, starting the integration outside the apsides allows us to sample complementary regions of phase space that would otherwise remain inaccessible.

To construct the PSS, the pairs $(x, \dot{x})$ are recorded at the instants when the trajectory crosses the plane $y = 0$ with $\dot{y} < 0$. In the numerical implementation, the crossing is detected by the change in sign of the variable $y$, and the corresponding values of $x$ and $\dot{x}$ are determined by linear interpolation.

Since the integration of the equations of motion is performed in the rotating frame, the determination of the initial velocity component \(\dot{y}_0\) depends on the position of the particle relative to the synchronous radius, defined as \(a_{\mathrm{sync}} = \lambda^{2/3} \), corresponding to the orbital distance at which the particle's period is equal to the central body's rotation period. Particles with \(a < a_{\mathrm{sync}}\) have \(n > \omega\), while those with \(a > a_{\mathrm{sync}}\) have \(n < \omega\), which determines the sign of the relative velocity in the rotating reference frame and must be taken into account in the construction of PSS.

\subsection{Lyapunov Characteristic Exponent}

In addition to PSS, we use the LCE ($\gamma$) as a tool for characterising and quantifying chaos. This parameter quantifies the average rate of divergence between trajectories that are initially close in phase space, a procedure that is detailed by \cite{benettin1980lyapunov}. In this work, this analysis is performed specifically using the Variational Equations method \citep[see][Chapter~5]{beutler2005methods}. Unlike approaches based solely on monitoring the Euclidean distance between neighbouring trajectories \citep{meyer1984regular, Murray1999}, this technique allows the average divergence rate to be extracted directly from the evolution of the space tangent to the reference trajectory \citep{beutler2004methods, lichtenberg2013regular}, generating computational cost savings.

To this end, we monitor the evolution of a deviation vector $\mathbf{w} = [\delta x, \delta y, \delta \dot{x}, \delta \dot{y}]^T$, responsible for measuring the total variation in the particle's state. The temporal evolution of this variation is governed by the numerical integration of an eight-dimensional system, which couples the original equations of motion to the variational equations linearised in matrix form \citep[see Chapter~5 of][]{lichtenberg2013regular}:
\begin{equation}
\dot{\mathbf{w}} = J(\mathbf{X})\,\mathbf{w},
\label{Eq_Variacional}
\end{equation}

\indent Here, $\dot{\mathbf{w}}$ represents the time derivative of the deviation vector, and $J(\mathbf{X})$ is the Jacobian of the system, responsible for propagating the deviation of all elements of the state $\mathbf{X}=(x,y,\dot{x},\dot{y})^T$:
\begin{equation}
J =
\begin{pmatrix}
0 & 0 & 1 & 0 \\
0 & 0 & 0 & 1 \\
J_{31} & J_{32} & 0 & 2\lambda \\
J_{41} & J_{42} & -2\lambda & 0
\end{pmatrix}.
\end{equation}

The elements $J_{ij}$ depend on the second derivatives of the gravitational potential $U(\mathrm{r})$ (Equation~\ref{Potencial}), and are expressed analytically as:
\begin{align}
J_{31} &= \lambda^2 - \frac{\partial F_x}{\partial x} = \lambda^2 + \sum_{i=1}^2 \mu_i \left( \frac{3(x-x_i)^2}{r_i^5} - \frac{1}{r_i^3} \right) \\
J_{32} &= - \frac{\partial F_x}{\partial y} = J_{41} = - \frac{\partial F_y}{\partial x} =  \sum_{i=1}^2 \mu_i \left( \frac{3y(x-x_i)}{r_i^5} \right) \\
J_{42} &= \lambda^2 - \frac{\partial F_y}{\partial y} =\lambda^2 + \sum_{i=1}^2 \mu_i \left( \frac{3y^2}{r_i^5} - \frac{1}{r_i^3} \right)
\end{align}
\indent In these expressions, $\mu_1 = 1-\mu$ and $\mu_2 = \mu$ represent the normalised masses of the spherical portion of the central body and the crater, located at $r_1=\sqrt{(x-x_1)^2+y^2}$ and $r_2=\sqrt{(x-x_2)^2+y^2}$, respectively.

The Lyapunov Exponent in an finite time ($\gamma$) is then defined by \citep{lichtenberg2013regular}:

\begin{equation}
\gamma(t) = \frac{1}{t} \ln \left( \frac{d(t)}{d_0} \right),
\end{equation}
\noindent where $d_0 = \|\mathbf{w}(0)\|$ is the initial separation in phase space, and $d(t) = \|\mathbf{w}(t)\|$ is the Euclidean norm of the deviation vector, which we monitor for divergence magnitude as a function of time. This metric integrates the contribution of all state variables \citep{beutler2004methods, lichtenberg2013regular}:
\begin{equation}
d(t) = \sqrt{\delta x^2 + \delta y^2 + \delta \dot{x}^2 + \delta \dot{y}^2}.
\end{equation}

Therefore, the numerical integration of Equation \eqref{Eq_Variacional} provides the temporal evolution of the deviation vector $\mathbf{w}(t)$. It is from the magnitude of this vector, expressed by its norm $d(t)$, that it becomes possible to calculate the exponential growth rate described by the LCE.

Since the LCE is formally defined in the asymptotic limit, its numerical estimate over a finite integration interval depends on the adopted integration time. For chaotic trajectories, the finite-time estimate $\gamma(t)$ is expected to converge towards a finite positive value, whereas for regular motion it approaches zero only asymptotically. In particular, for regular quasi-periodic motion, the conventional Lyapunov estimator decreases approximately as $\ln(t)/t$ \citep{cincotta2000simple, cincotta2003phase}. Consequently, a finite-time calculation may return a small but non-zero value of $\gamma$ even for a regular trajectory, and increasingly long integrations may be required to distinguish regular, weakly chaotic, and sticky motion.

Accordingly, the system is classified as regular when $\gamma(t)$ tends towards zero and as chaotic when it converges towards a finite positive value. The corresponding Lyapunov time, $T_L = 1/\gamma$, provides the characteristic timescale over which an initial separation in phase space grows by a factor of~ $\overline{e} \approx 2.718$ \citep{benettin1980lyapunov}. For chaotic trajectories whose LCE has reached convergence, $T_L$ provides a characteristic predictability timescale, with smaller values indicating faster exponential divergence. For regular or weakly chaotic trajectories for which $\gamma(t)$ has not reached asymptotic convergence, $T_L$ should instead be interpreted as a finite-time dynamical indicator and used comparatively rather than as an absolute long-term stability timescale.

\section{Exploring the nominal case} \label{Secao3}

In this section, we will define our nominal case, i.e., the case in which the physical parameters are based on Máni's data by \cite{Rommel_2023}, showing how these parameters were considered and the results obtained for the description of stability in this system.

\subsection{Physical parameters for Máni}

The physical characteristics of TNO Máni estimated by \cite{Rommel_2023} are presented in Table \ref{tab:ocultacoes_MS4}. For the stability analysis performed in this work, we adopted the global density $\rho = 720$~kg/m$^3$ and the rotation period $T_{\mathrm{rot}} = 7.44$~h. Although \cite{Rommel_2023} suggests that  Máni can be approximated by a Maclaurin ellipsoid with equatorial axes $A = B \approx 401.37$~km and polar axis $C = 234$~km, the model adopted in this work simplifies the central body to a sphere with an equivalent volume radius, $R_c \approx 335$~km. This simplification allows us to focus on the gravitational perturbation caused by the topography, specifically the reported large crater. 

The crater is modeled as a spherical structure with a radius $r_c~=~45$~km, equal to its estimated depth. Assuming a uniform density, the mass of the central body is $M_0 = 1.13 \times 10^{20}$~kg, and the mass removed to form the crater is $m = -2.75 \times 10^{17}$~kg. These values lead to the dimensionless mass parameter $\overline{\mu} \approx 2 \times 10^{-3}$, which represents the ratio between the mass of the crater and the central body. Finally, the system's rotation rate, $\lambda$, is derived from the rotation period $T_{\mathrm{rot}}$, resulting in $\lambda \approx 0.52$.

\begin{table*}
\centering
\caption{Summary of the main results of stellar occultations for Máni \protect\citep{Rommel_2023}}
\label{tab:ocultacoes_MS4}
%\resizebox{\columnwidth}{!}{% % Isso força a tabela a caber na coluna sem erros
\begin{tabular}{|l|p{6cm}|}
\hline
\textbf{Parameter} & \textbf{Estimate} \\ \hline
Projected major axis ($A$) & $412 \pm 10$ km \\ \hline
Projected minor axis ($B$) & $385 \pm 17$ km \\ \hline
Estimated polar radius ($C$) & Minimum 234~km and maximum 387~km \\ \hline
Diameter equivalent to projected area & $796 \pm 24$ km \\ \hline
Rotation period \citep{thirouin2013study} & $7.44$~h or $10.44$~h \\ \hline
Global density & $0.72~g.cm^{-3}$ or $0.36~g.cm^{-3}$ \\ \hline
Topography & Depression of 11 km; Peak of 25 km; Crater of 322 km and 45 km deep \\ \hline
\end{tabular}%
%}
\end{table*}

With the nominal parameters  based on Maní, the next section addresses the process of investigating stability in this system using the PSS technique.

\subsection{Application of the Poincaré Surfaces of Section technique} \label{Secao3.2}

\begin{figure*}
    \centering
    % Primeira Subfigura (Esquerda)
    \begin{subfigure}{0.48\textwidth}
        \centering
        \includegraphics[width=\linewidth]{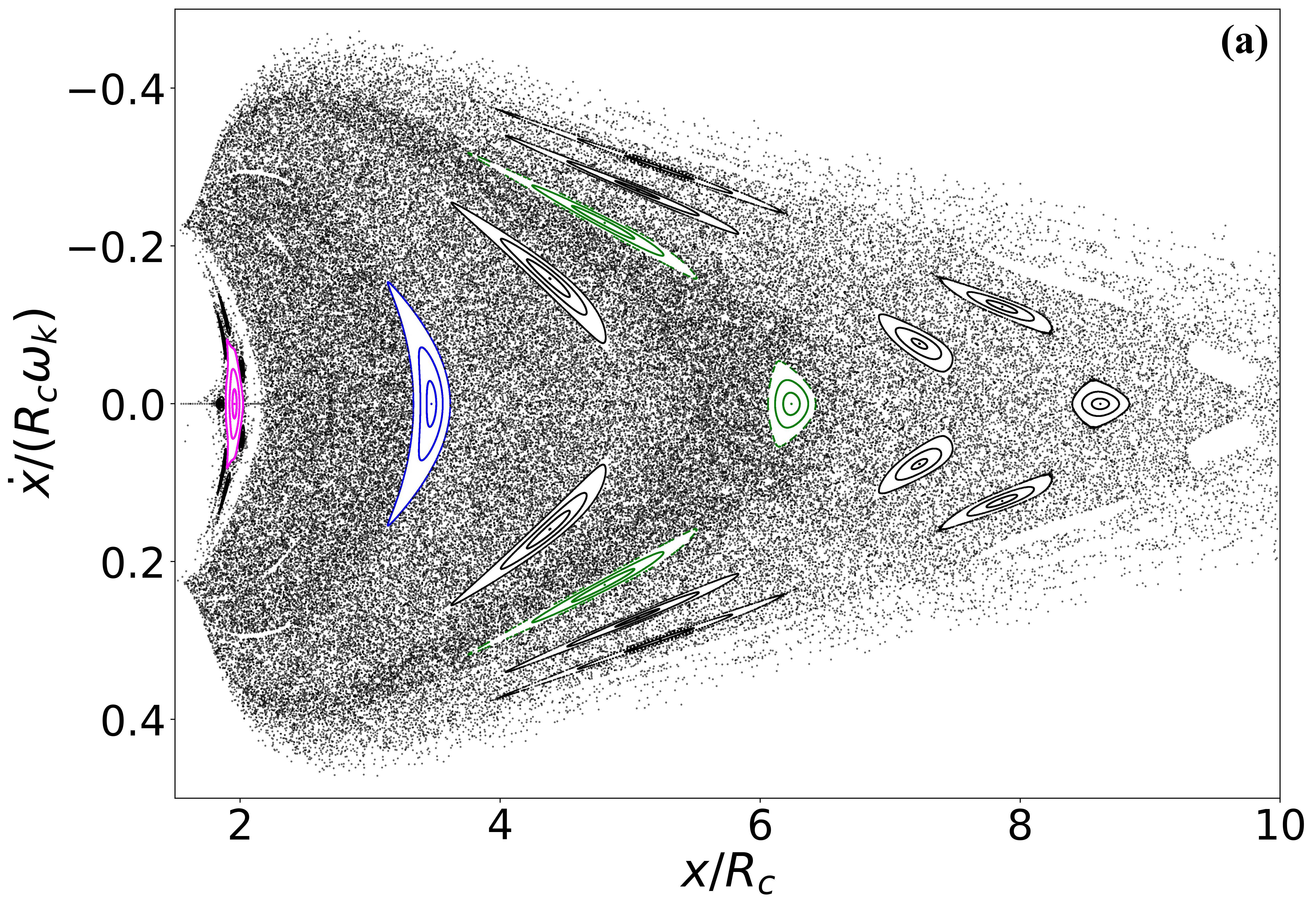}
%        \caption{$C_J = 1{,}800 \, R_c^2 \omega_k^2$.}
        \label{SSP1.800}
    \end{subfigure}
    \hfill % Espaçamento horizontal entre as figuras
    % Segunda Subfigura (Direita)
    \begin{subfigure}{0.48\textwidth}
        \centering
        \includegraphics[width=\linewidth]{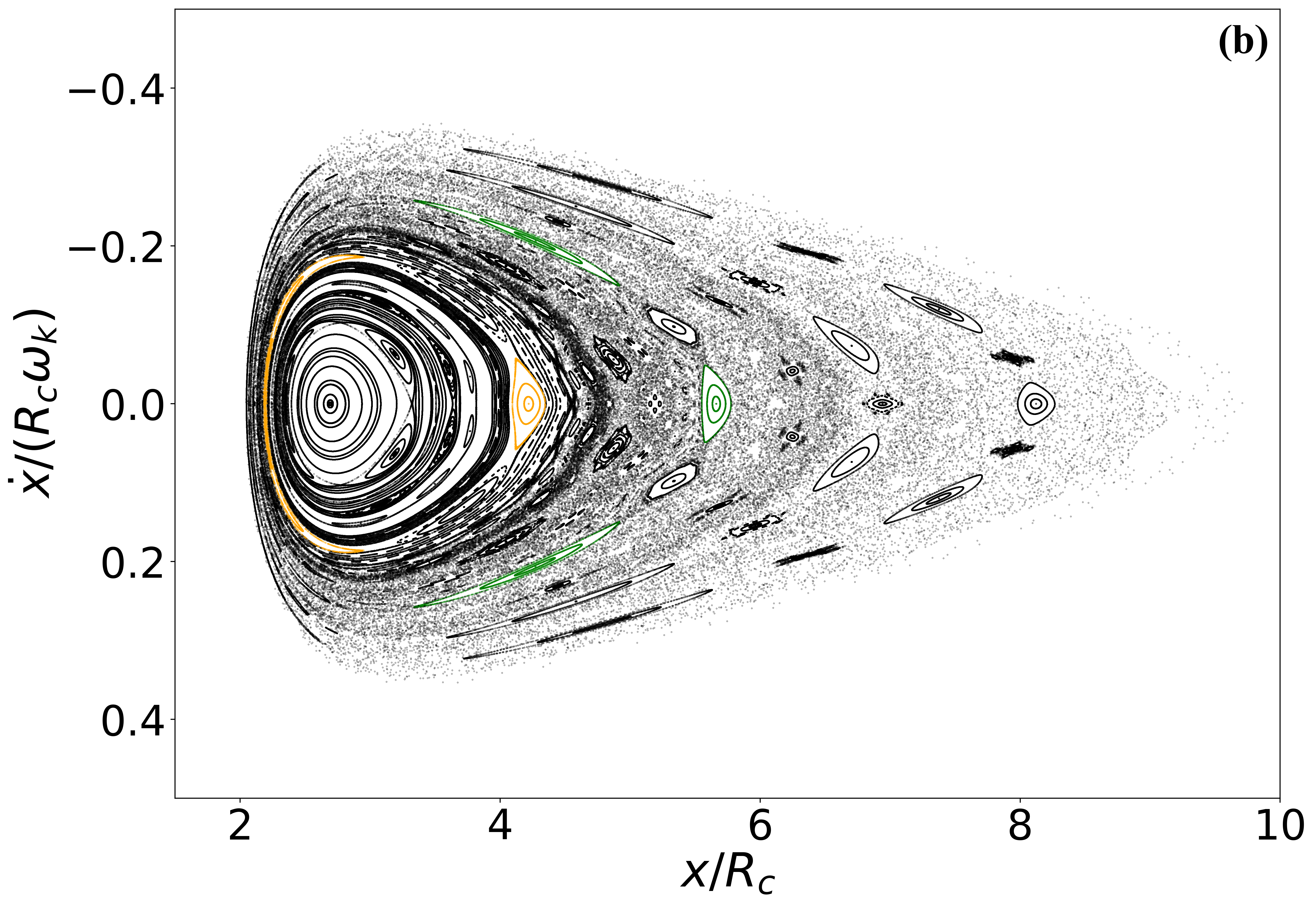}
%        \caption{$C_J = 2{,}167 \, R_c^2 \omega_k^2$.}
        \label{SSP2.167}
    \end{subfigure}
    
    \caption{PSS for the case of the central body with $\overline{\mu}=2 \times 10^{-3}$ and $\lambda=0.52$, for (a) $C_J=1.890~R_c^2 \omega_k^2$ and (b) $C_J=2.085~R_c^2 \omega_k^2$. In panel (a), the islands of stability are highlighted in colour: pink represents a structure near the $1\!:\!1$ SOR, blue the $2\!:\!1$ SOR, and green the $4\!:\!1$ SOR. The black structures represent fourth-order or higher SORs, and the dotted region indicates the chaotic domain. In (b), for $C_J = 2.085~R_c^2 \omega_k^2$, the system exhibits structures associated with trajectories of lower eccentricities compared to (a). Note that among the highlighted structures in (a), only the green one ($4\!:\!1$ SOR) remains, while the $3\!:\!1$ SOR (orange) is now represented for this value of $C_J$. Concurrently, the expansion of the first-type orbit structure is observed, which suppresses the remaining resonant islands as they shift and eventually collapse into its core with further increments in $C_J$.}
    \label{fig:painel_poincare}
\end{figure*}

The PSS is constructed from a regular grid of 40 initial conditions distributed uniformly in the interval $x_0 \in [1.0; 7.0]\,R_c$. This initial configuration provides an overview of the organisation of structures in phase space. Based on this mapping, new particles are added punctually in regions of interest to refine and delineate the dynamic structures present in greater detail. 

Figure~\ref{fig:painel_poincare} (a) shows the PSS for \(C_J = 1.890\,R_c^2 \omega_k^2\). This section was generated using 93 particles: the initial standard grid was supplemented with 53 additional particles to refine the resolution of the main resonant structures. Following the methodology previously described, most particles were initialized at the apocentre of their osculating orbit, with \(y_0=0\) and \(\dot{x}_0=0\), while \(\dot{y}_0\) was determined from the Jacobi constant. For initial conditions on the \(x\)-axis, \(\dot{x}_0\) corresponds to the radial component of the velocity. Therefore, adopting \(\dot{x}_0\neq0\) implies that the particle is not initially located at an apsis of its osculating orbit. Some SOR islands were found only for such initial conditions. A detailed analysis of these families lies outside the scope of the present study and will be addressed in future work.

%Figure~\ref{fig:painel_poincare} (a) shows the PSS for $C_J = 1.890\,R_c^2 \omega_k^2$. This section was generated using 93 particles: the initial standard grid was supplemented with 53 additional particles to refine the resolution of the main resonant structures. Following the methodology previously described, the majority of the particles were launched from the apocentre with $\dot{x}_0 = 0$, while $\dot{y}_0$ was determined by the Jacobi constant. It is worth noting, however, that certain SORs islands were obtained by assuming initial conditions with $\dot{x}_0 \neq 0$; a detailed analysis of these structures lies outside the scope of this study and will be presented in a forthcoming paper in preparation.

Analysis of Figure \ref{fig:painel_poincare} (a) reveals a region of instability in the immediate vicinity of the central body, where trajectories result in collision, defined either by the particle reaching $r \leq R_c$ or when its orbital pericentre lies within the physical radius of the body, or ejection ($r \geq 100\,R_c$ or $e \geq 1$) on timescales shorter than $1\,T_{\mathrm{rot}}$. In the plane $(x, \dot{x})$, this region extends along the axis $\dot{x} = 0$ to $x \approx 1.8\,R_c$, characterised by points distributed horizontally at $\dot{x}=0$. Chaotic regions are delimited by the irregular distribution of points in phase space. This sensitivity to initial conditions implies that the future state of a trajectory is unpredictable, preventing the determination of the behaviour of particles that start from almost identical states \citep{lecar2001chaos}. Consequently, the long-term fate of a particle in this region cannot be determined,  it may collide, be ejected, or remain in an unpredictable orbit.

Each island of stability is composed of a set of closed curves representing quasi-periodic orbits. Although these trajectories remain confined to a surface in phase space, they are not strictly periodic; the orbit returns arbitrarily close to previous states, but the dynamics never repeat exactly at fixed time intervals. These curves circle around a central fixed point that corresponds to a periodic orbit, where the particle returns to the exact same position in phase space at regular intervals, resulting in predictable dynamics. As the curves approach this centre, their size decreases, eventually converging to the central periodic orbit. The boundary that delimits these islands, separating them from the chaotic region or from adjacent islands, is defined as the SOR separatrix.

\begin{figure}
    \centering
    \includegraphics[width=1.\linewidth]{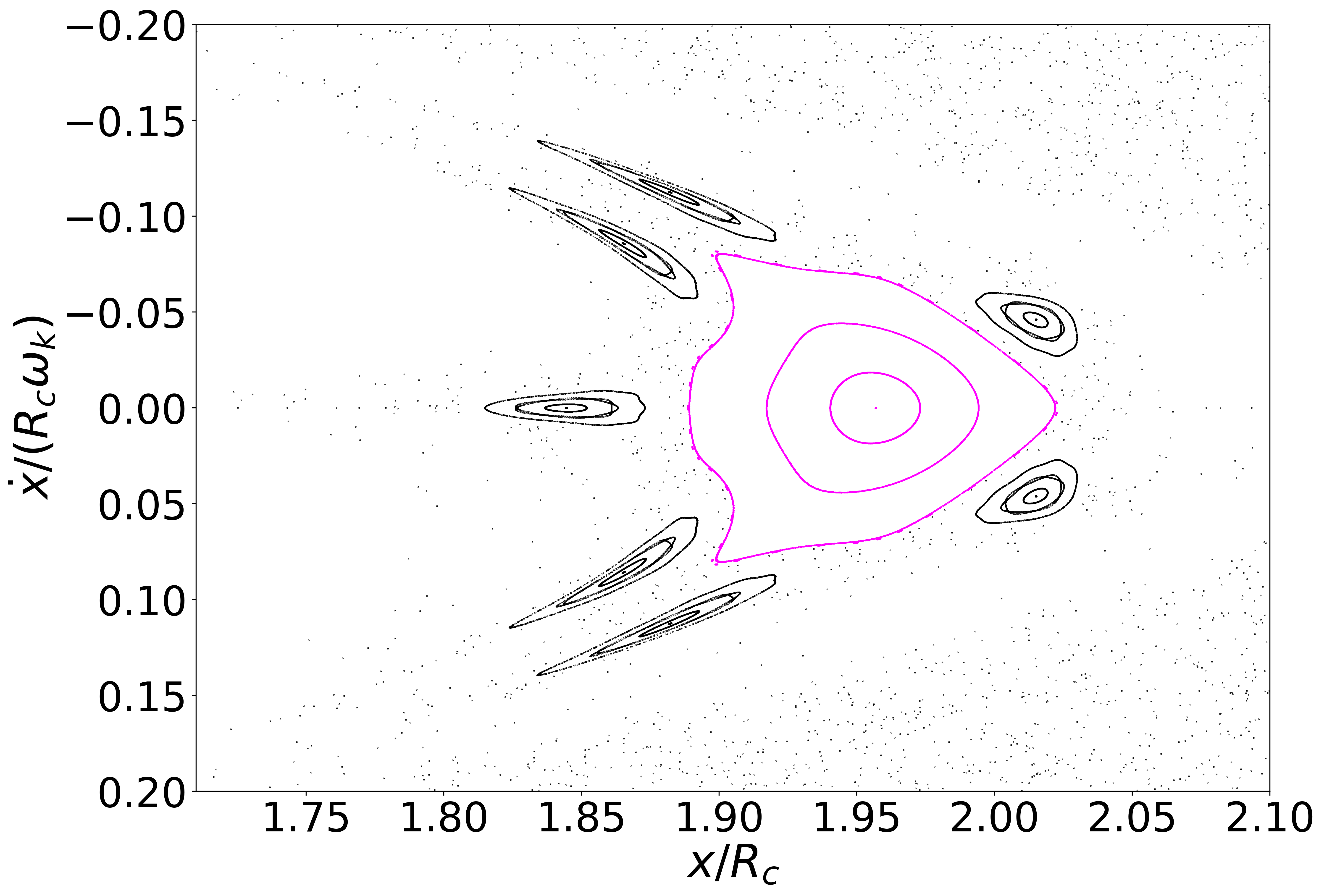}
    \caption{Detailed view of the phase space region near the $1\!:\!1$ SOR (pink structure) for $C_J = 1.890\,R_c^2 \omega_k^2$. This zoom reveals the presence of higher-order resonant structures in the immediate vicinity of the $1\!:\!1$ SOR island.}
    \label{fig:zoom_1_1}
\end{figure}

The PSS in Figure \ref{fig:painel_poincare} (a) displays structures of islands associated with different SORs in phase space, classified according to the criteria of \cite{Murray1999} and \cite{morbidelli2002modern}. The innermost islands, highlighted in pink and blue, appear as isolated structures formed by a single island; specifically, the pink structure corresponds to the $1\!:\!1$ SOR, which occurs when the orbital period of a particle is near or matches the rotation period of the central body, while the blue structure represents the $2\!:\!1$ SOR. A closer inspection of the region surrounding this $1\!:\!1$ SOR domain (see Figure \ref{fig:zoom_1_1}) reveals that this stability island is flanked by smaller structures associated with higher-order SORs. The structure highlighted in green is formed by three islands and is identified as the third-order $4\!:\!1$ SOR. The remaining structures shown in black correspond to fourth-order or higher SORs; however, additional structures associated with the $3\!:\!1$ and $4\!:\!1$ SORs also appear when the initial $\dot{x}_0 \neq 0$ of the particles is assumed, which lies beyond the scope of this study.

Figure~\ref{fig:painel_poincare}~(b) shows the PSS for a larger Jacobi constant value, $C_J = 2.085\,R_c^2 \omega_k^2$. This section was generated using 138 particles, where the initial standard grid was supplemented with 98 additional particles to refine the resolution of the main resonant structures. In this scenario, the stability islands associated with the remaining SORs are located slightly closer to the central body. Consequently, a significant reduction in points associated with chaotic trajectories is observed when compared to Figure~\ref{fig:painel_poincare}~(a). 

Among the highlighted structures, the second-order $3\!:\!1$ SOR is represented in orange, composed of a pair of islands located at the $\dot{x}_0 = 0$ axis. With further increments in the value of $C_J$, these resonant structures tend to shift progressively toward the center of the first-kind orbit domain until they eventually collapse into it. This shifting and subsequent collapse occur because higher values of the Jacobi constant are associated with lower-eccentricity trajectories; this dynamically suppresses the resonances while favoring the expansion of the stable region linked to the first-kind orbit (for details, see \citealt{Madeira2022, Ribeiro2023}). Consequently, this main regular domain expands to the point of covering and absorbing the islands associated with the SORs.

%Following \cite{winter1997resonance}, the construction of the $a\times e$ grid from the PSS was performed for the nominal case. Considering the interval $C_J \in [1{.}627, 2{.}223] \, R_c^2 \omega_k^2$, the central points (referring to periodic orbits) associated with SORs were extracted, as well as the inner limits, $(x_0, \dot{x}_0)_{\text{int}}$, and outer limits, $(x_0, \dot{x}_0)_{\text{ext}}$, of the islands of stability. These points delimit the outermost quasi-periodic orbits associated with each SOR studied, located at the separatrix of the structure. This procedure allowed us to map the width of the resonances with eccentricity, identifying the boundaries where the motion ceases to be regular and becomes chaotic or subject to adjacent SORs.

\begin{figure*}
    \centering
    \includegraphics[width=1.0\linewidth]{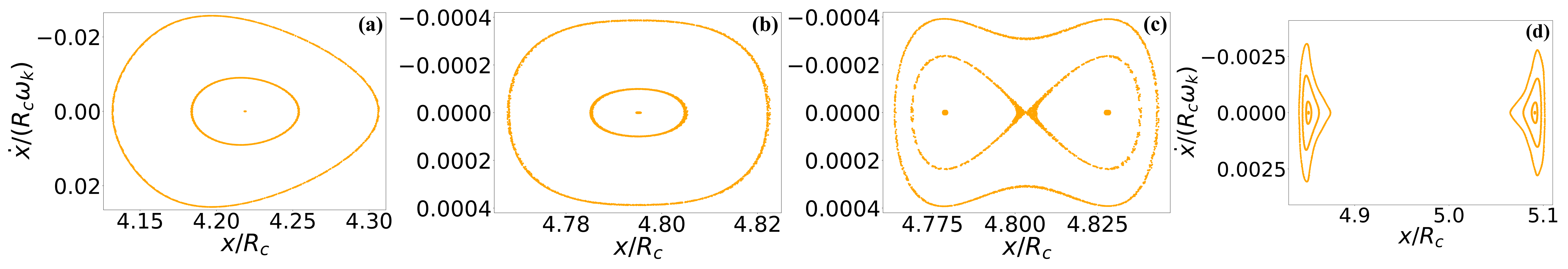}
    \caption{Successive PSS zooms illustrating the period-doubling bifurcation of the second-order $3\!:\!1$ SOR for: (a) $C_J = 2.085$, (b) $1.9328$, (c) $1.9300$, and (d) $1.869 \, R_c^2 \omega_k^2$. The sequence tracks the structural transition from a single central stability island at $\dot{x}_0 = 0$ (panels a and b) to a bifurcated state resulting in an effective $6\!:\!2$ SOR (panels c and d), with the critical threshold taking place within the interval $C_J \in (1.9300, 1.9328) \, R_c^2 \omega_k^2$.}
    \label{fig:bifurcacao_pss}
\end{figure*}

By tracking the evolution of the second-order $3\!:\!1$ SOR across different energy levels, a remarkable dynamical feature is identified. Within the interval $C_J \in (1.9300, 1.9328)\,R_c^2\omega_k^2$, a period-doubling bifurcation takes place. Bifurcations of resonant families have also been identified in the two preceding studies of this series on the dynamics around non-spherically symmetric bodies, of which the present work represents a continuation, although under different dynamical systems. In the mass-anomaly model of \cite{Madeira2022}, the $4\!:\!1$ SOR (denoted $1\!:\!4$ in their convention) bifurcates after a critical eccentricity is reached as $C_J$ decreases, with a single stable fixed point splitting into two independent resonant families associated with asymmetric libration. In the prolate-body model of \cite{Ribeiro2023}, a bifurcation of the $6\!:\!2$ SOR (denoted $2\!:\!6$ in their convention) appears for a configuration with a reduced rotation rate, producing additional phase-space crossings and a chaotic separatrix that restricts the stable extent of the resonance. In the present crater model, the $3\!:\!1$ bifurcation likewise develops along the evolution of the resonant family with $C_J$, but occurs exactly at $\dot{x}_0=0$. Thus, although the resonances and the underlying dynamical configurations differ among the three models, the results consistently show that resonant families around non-spherically symmetric bodies can undergo substantial topological reorganisations that modify the surrounding regular and chaotic domains.

%By tracking the evolution of the second-order $3\!:\!1$ SOR across different energy levels, a remarkable dynamical feature is identified. Within the interval $C_J \in (1.9300, 1.9328) \, R_c^2 \omega_k^2$, a period-doubling bifurcation takes place. This type of bifurcation and its consequent period-doubling effect are known to occur in certain retrograde resonance regimes \citep[see][]{carita2023retrograde}. Furthermore, similar structural transitions for this specific resonance have been reported by \cite{Madeira2022} and \cite{Ribeiro2023}. However, a key distinction lies in the phase space topology: in both previous studies, the bifurcation occurs at $\dot{x}_0 \neq 0$, whereas in our model it takes place exactly at $\dot{x}_0 = 0$.

To illustrate this mechanism from both statistical and geometric perspectives, two complementary panels are introduced. Figure \ref{fig:bifurcacao_pss} displays successive zooms of the PSS around the $3\!:\!1$ SOR structure, showing the evolution of the phase space. For higher values of the Jacobi constant, such as $C_J = 2.085 \, R_c^2 \omega_k^2$ (Figure \ref{fig:bifurcacao_pss}a) and $C_J = 1.9328 \, R_c^2 \omega_k^2$ (Figure \ref{fig:bifurcacao_pss}b), the resonance is characterised by standard islands centered at the core of the structure. Below the critical threshold, the bifurcation splits the central fixed points into a pair of periodic and quasi-periodic orbits (Figure \ref{fig:bifurcacao_pss}c and d). This phenomenon results in a period-doubling effect, where the resonance effectively transitions into a $6\!:\!2$ state.

\begin{figure*}
    \centering
    \includegraphics[width=1\linewidth]{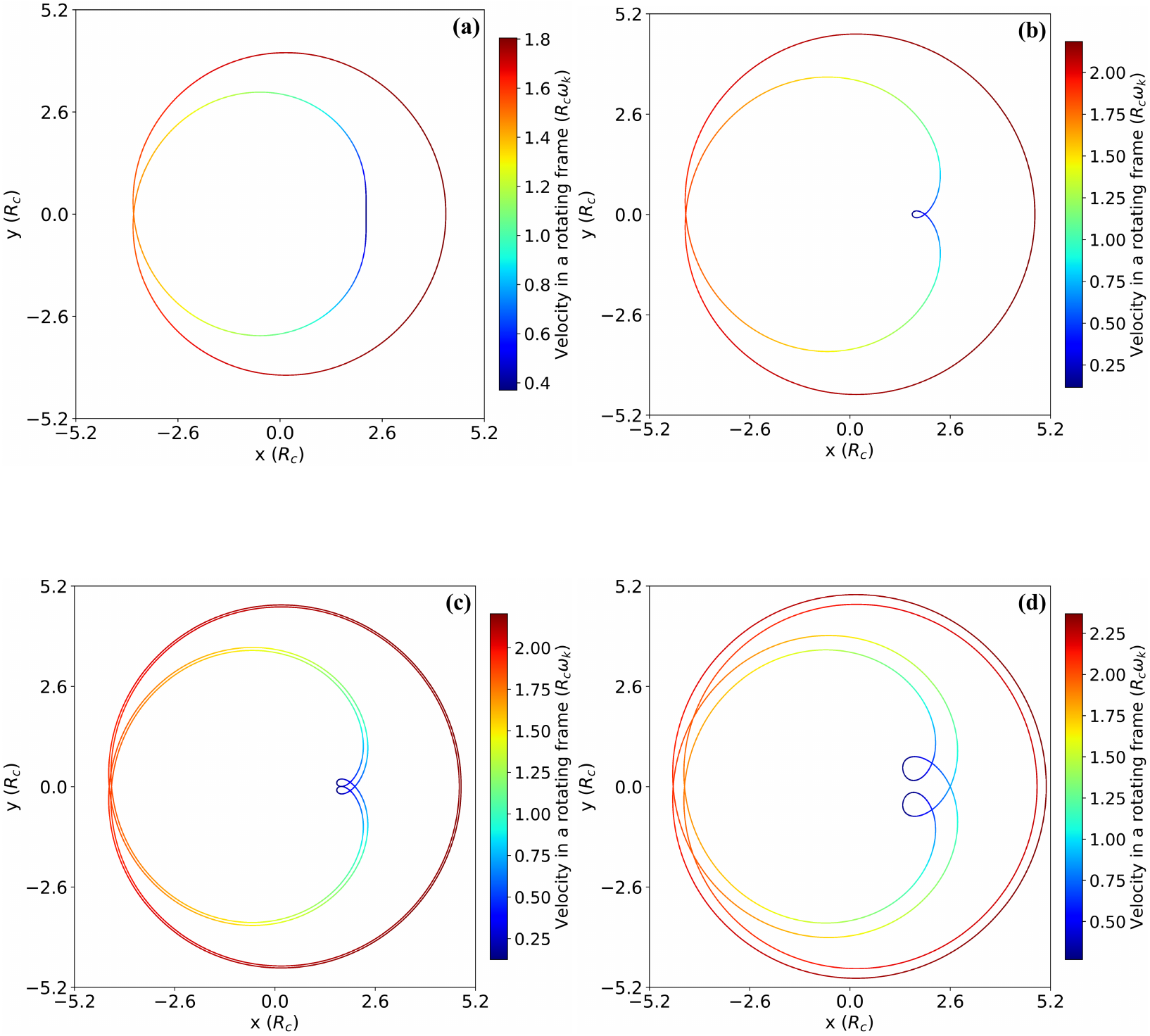}
    \caption{Periodic orbits associated with the center of the $3\!:\!1$ (and bifurcated $6\!:\!2$) SOR within the rotating reference frame for: (a) $C_J = 2.085$, (b) $1.9328$, (c) $1.9300$, and (d) $1.869 \, R_c^2 \omega_k^2$. The sequence shows the structural evolution and the appearance of internal loops after the period-doubling bifurcation, maintaining reflection symmetry with respect to $y = 0$. The colorbar indicates the magnitude of the velocity vector in the rotating frame.}
    \label{fig:trajetorias_girante}
\end{figure*}

Complementarily, Figure \ref{fig:trajetorias_girante} presents the physical manifestation of this transition by mapping the periodic orbits at the center of these structures within the rotating reference frame. For $C_J = 2.085 \, R_c^2 \omega_k^2$ and $C_J = 1.9328 \, R_c^2 \omega_k^2$, the particle follows a single regular closed trajectory. Analising the bifurcation ($C_J \leq 1.9300 \, R_c^2 \omega_k^2$), the orbit undergoes a geometric duplication. Notably, unlike the asymmetric trajectories found by \cite{Madeira2022}, our system yields periodic orbits in the rotating frame that are perfectly symmetric with respect to $y=0$. The colorbar in Figure \ref{fig:trajetorias_girante} indicates the magnitude of the velocity vector in the rotating frame, defined as $v = \sqrt{\dot{x}^2 + \dot{y}^2}$ along the path, revealing that the particle's velocity drops significantly near these turning points and loops.

%Notably, unlike the asymmetric trajectories found by \cite{Madeira2022}, our system yields periodic orbits in the rotating frame that are perfectly symmetric with respect to $y=0$. The colorbar in Figure \ref{fig:trajetorias_girante} indicates the magnitude of the velocity vector in the rotating frame, defined as $v = \sqrt{\dot{x}^2 + \dot{y}^2}$, along the path. This reveals that the particle's velocity drops significantly near these turning points and loops. This detailed orbit tracking directly accounts for the dual-branch geometry observed for high eccentricities in the subsequent grid analysis.}

\begin{figure}
        \centering
        \includegraphics[width=1.\linewidth]{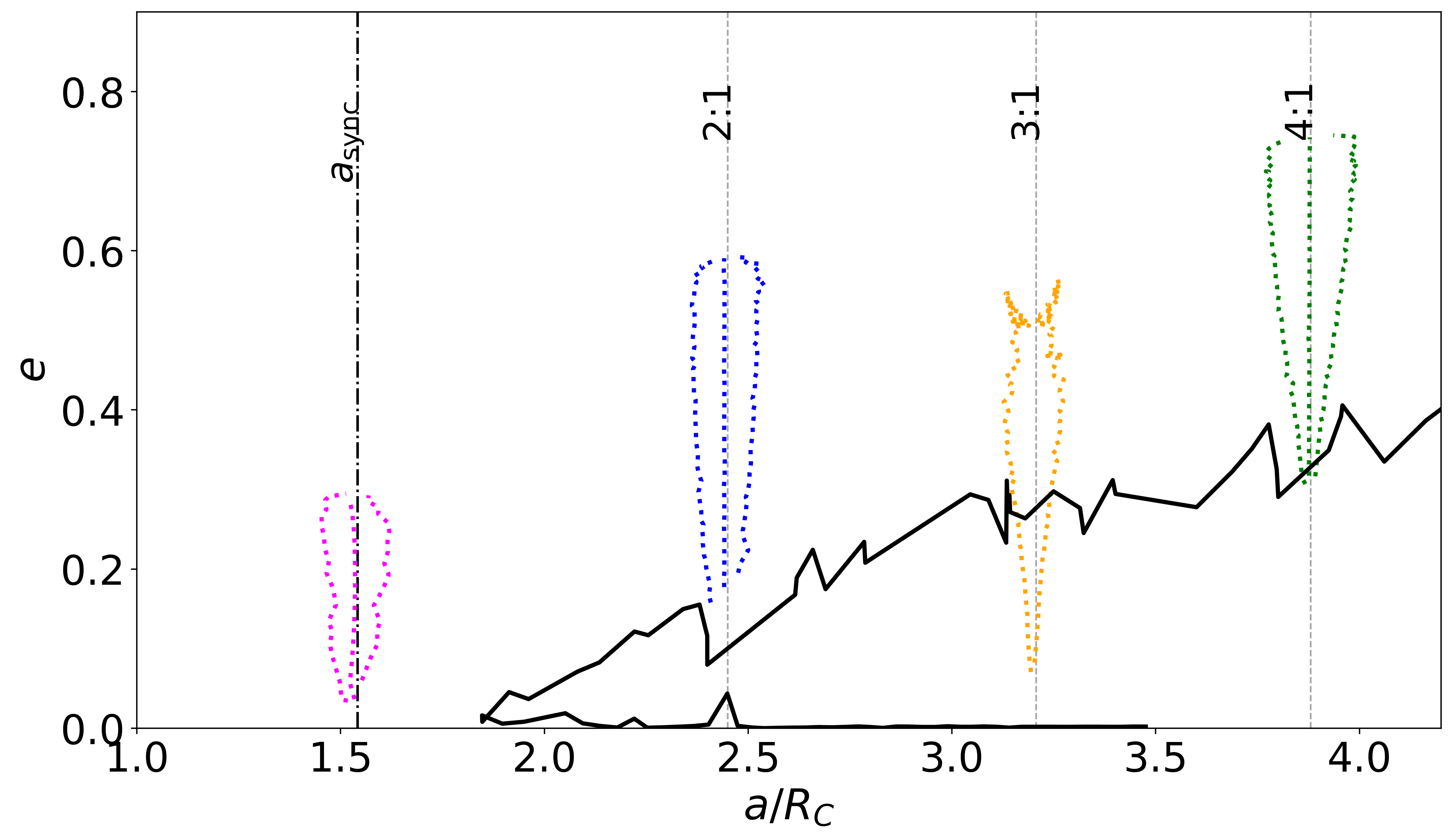}
        \caption{Grid $a \times e$ with the positions of SORs calculated theoretically, and with the width and their respective centres calculated numerically from data extracted from PSSs. The black outline delimits the stable region associated with the boundaries of first-type orbits. Coloured dashed lines correspond to the limits of the SORs identified in Fig.~\ref{fig:painel_poincare}, mapped here by tracking their evolution across several PSS analyses. Theoretical SOR positions are indicated by black dashed lines.}
        \label{gradeapore}
    \end{figure}

Following the methodology of \cite{winter1997resonance}, the boundaries of these stability islands—including the central periodic orbits, inner limits $(x_0, \dot{x}_0)_{\text{int}}$, and outer limits $(x_0, \dot{x}_0)_{\text{ext}}$ located at the separatrices—were extracted from the PSSs across the interval $C_J \in [1{.}627, 2{.}223] \, R_c^2 \omega_k^2$. These pairs were transformed from the rotational reference frame to the inertial frame components $(X_0, \dot{X}_0)$ to calculate the semi-major axis ($a$) and eccentricity ($e$) via the standard equations from \cite{Murray1999}. This procedure allowed us to map the resonance widths with eccentricity, identifying the boundaries where the motion ceases to be regular and defining the global regular domains shown in Figure \ref{gradeapore}.

%The points extracted from the PSSs were transformed from the rotational reference frame to the inertial reference frame. From the components $(X_0, \dot{X}_0)$ in the inertial reference frame, we calculated the semi-major axis ($a$) and eccentricity ($e$) using the equations from \cite{Murray1999}. This procedure allowed us to associate each pair $(x_0, \dot{x}_0)$ of the section with a corresponding pair $(a, e)$, whose results are shown in Figure \ref{gradeapore}.

As displayed in Figure \ref{gradeapore}, the colored dots represent the stable regions of the mapped SORs. Notably, although the structure associated with the $1\!:\!1$ SOR (pink region) is located very close to the central body, it maintains its stability profile across low eccentricities, as bounded by the numerical PSS integration limits. At higher eccentricities, the $3\!:\!1$ SOR structure develops a distinct split-path geometry, widening into two prominent upper tips for $e > 0.5$. This split directly reflects the period-doubling bifurcation into the $6\!:\!2$ SOR state previously discussed and mapped in Figures \ref{fig:bifurcacao_pss} and \ref{fig:trajetorias_girante}.

Additionally, the region outlined by the black line in Figure \ref{gradeapore} marks the boundaries of the first-type orbit domain, which is inherently characterised by low eccentricities and emerges in the PSS for $C_J \geq 1.965 \, R_c^2 \omega_k^2$. Within this enclosed curve, the trajectories remain highly stable. The lower boundary of this curve provides a quantitative measure of the minimum possible eccentricity within the first-type orbit domain, a threshold that systematically decreases as the semi-major axis increases; for instance, at $a = 2.052 \, R_c$, the boundary establishes a minimum eccentricity of $e = 0.022$.

Table \ref{tab:ressonancias_completa} summarises the theoretical positions and the numerical deviations for the main resonances. A relevant detail is that the structures associated with the $4\!:\!1$, $3\!:\!1$, $2\!:\!1$, and $1\!:\!1$ SORs exhibit a very slight shift to the right of the theoretical positions calculated via Equation \ref{spinorbita}. This discrepancy is quantified by the mean percentage deviation, $D_p$, which reaches a maximum of $2.82\%$ for the $1\!:\!1$ SOR and progressively decreases for structures located further from the central body. This trend occurs because Equation \ref{spinorbita} is derived from an analytical expansion that considers only the first-order term of the crater's perturbation. Consequently, at closer distances, the contribution of higher-order expansion terms neglected in the theoretical model becomes more pronounced, leading to the observed discrepancy. Conversely, at larger orbital radii, these terms decrease, resulting in a better alignment with the theoretical prediction.

%There is also a region outlined in a black line that is associated with the boundaries of the first-type orbit domain, which by definition is characterised by low eccentricities (found in the PSS for $C_J \geq 1.965~{R_c}^2{\omega_k}^2$). Within the area delimited by this curve, the motion remains stable. Notably, the lower boundary of this curve provides a measure of the minimum possible eccentricity within the first-type orbit domain, a threshold that decreases as the semi-major axis increases. For example, at a semi-major axis of $a = 2.052~R_c$, the minimum eccentricity of this boundary is $e = 0.022$.

\begin{table}
\centering
\caption{Theoretical positions and mean percentage deviation of the data obtained numerically for each SOR.}
\label{tab:ressonancias_completa}
\setlength{\tabcolsep}{4pt} % Compacta o espaço entre as colunas
\small % Define um tamanho de fonte fixo e elegante para tabelas
\begin{tabular}{|c|c|c|}
\hline
\textbf{SOR} & \textbf{Theoretical position $a$ ($R_c$)} & \textbf{Deviation ($D_p$)} \\ \hline
$1\!:\!1$ & 1.542 & 2.82\% \\ \hline
$2\!:\!1$ & 2.444 & 1.85\% \\ \hline
$3\!:\!1$ & 2.203 & 1.57\% \\ \hline
$4\!:\!1$ & 3.207 & 1.34\% \\ \hline
\end{tabular}
\end{table}

%Table \ref{tab:ressonancias_completa} summarises the theoretical positions and the numerical deviations for the main resonances. A relevant detail is that the structures associated with the $4\!:\!1$, $3\!:\!1$, $2\!:\!1$, and $1\!:\!1$ SORs exhibit a very slight shift to the right of the theoretical positions calculated via Equation \ref{spinorbita}. This discrepancy is quantified by the mean percentage deviation, $D_p$, which reaches a maximum of $2.82\%$ for the $1\!:\!1$ SOR and progressively decreases for structures located further from the central body. This trend occurs because Equation \ref{spinorbita} is derived from an analytical expansion that considers only the first-order term of the crater's perturbation. Consequently, at closer distances, the contribution of higher-order expansion terms neglected in the theoretical model becomes more pronounced, leading to the observed discrepancy. Conversely, at larger orbital radii, these terms decrease, resulting in a better alignment with the theoretical prediction.

\subsection{Survival maps} \label{sectionMapa}

In this technique, the dynamic structure of the system is investigated through the distribution of particles on a grid of initial conditions in the $(a, e)$ plane. For the nominal case, using parameters based on estimated data for Máni, we numerically integrated 160,000 particles arranged in a uniform mesh of $400 \times 400$ points, covering the domain $a \in [1.0; 4.2] \, R_c$ and $e \in [0.0001; 0.9]$. Regarding the remaining orbital elements, since our study is formulated entirely within a two-dimensional (2D) framework, all particles were initialised with a geometric equivalence of $I_0 = \Omega_0 = 0^\circ$, where $I_0$ represents the initial orbital inclination and $\Omega_0$ is the longitude of the ascending node. Additionally, they were assigned an argument of pericentre of $\omega_0 = 180^\circ$, which defines the orientation of the orbital ellipse in the study plane, and a true anomaly of $f_0 = 180^\circ$, which specifies the initial position of the particle along its orbit (in this case, apocentre); this configuration corresponds to a clockwise trajectory starting from the apocentre at $x_0 > 0$. The trajectories were calculated using the RK45 method, integrating the equations of motion (Equations \ref{eq:mov_x} and \ref{eq:mov_y}). 

The stability of each orbit is evaluated from its survival time and final state. Particles are classified as finite-time survivors if they remain in the system throughout the integration interval of $10^4\,T_{\mathrm{rot}}$. Particles that do not survive up to this limit are classified as collisions or ejections according to the distance and eccentricity criteria detailed in Section~\ref{Secao3.2}.

In the present work, the integration interval of $10^4\,T_{\mathrm{rot}}$ follows the timescale adopted in the previous studies of this series \citep{Madeira2022, Ribeiro2023}, with time expressed in units of the central body's rotation period. Survival over this interval is interpreted as a finite-time stability criterion rather than as evidence of indefinite stability.

\begin{figure}
    \centering
    % Primeira Subfigura (Esquerda)
    \begin{subfigure}{0.48\textwidth}
        \centering
        \includegraphics[width=\linewidth]{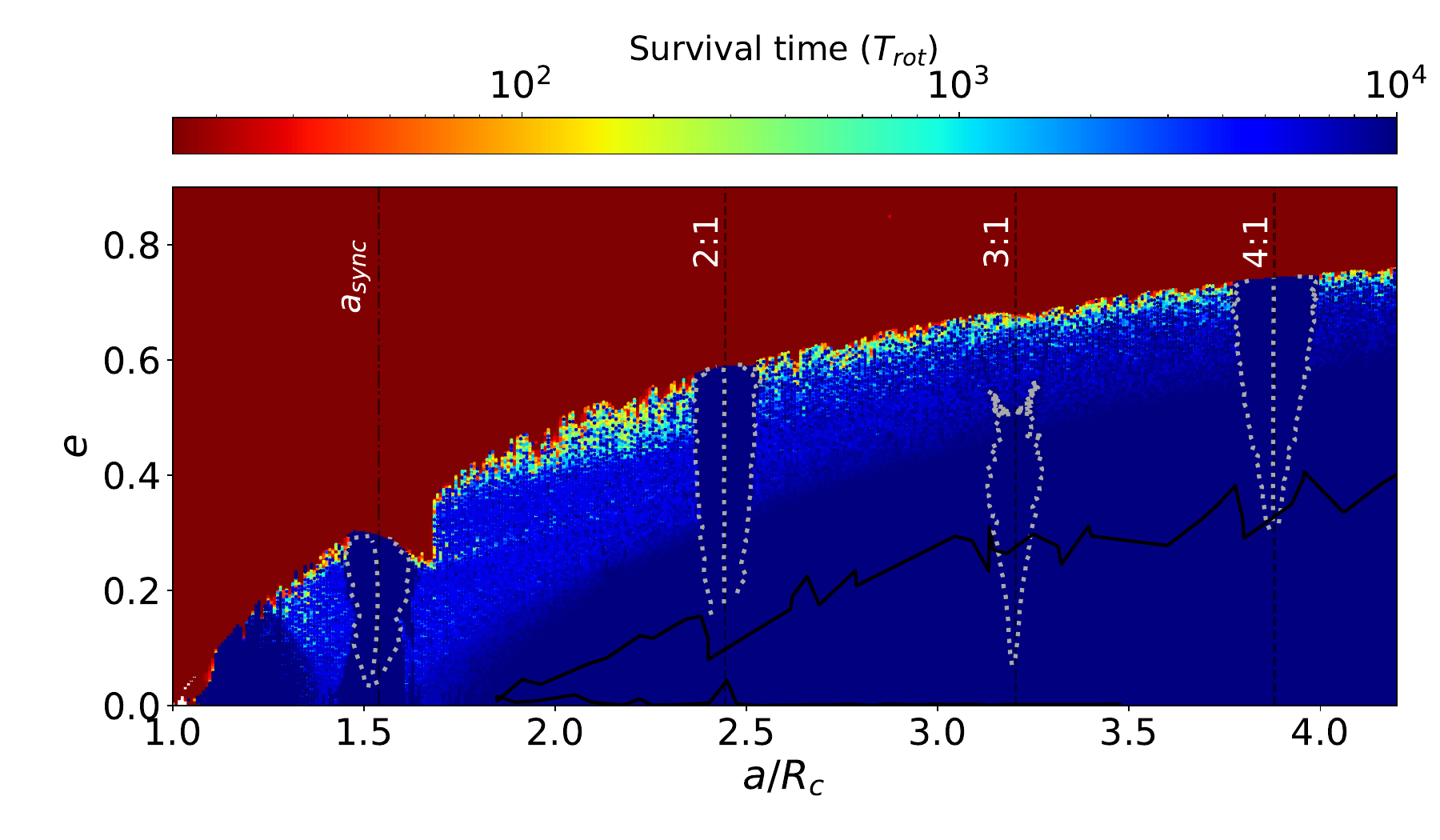}
%        \caption{$C_J = 1{,}800 \, R_c^2 \omega_k^2$.}
        \label{grade_ae_0.002_periodo_sob_COMBINADO}
    \end{subfigure}
    \hfill % Espaçamento horizontal entre as figuras
    % Segunda Subfigura (Direita)
    \begin{subfigure}{0.48\textwidth}
        \centering
        \includegraphics[width=\linewidth]{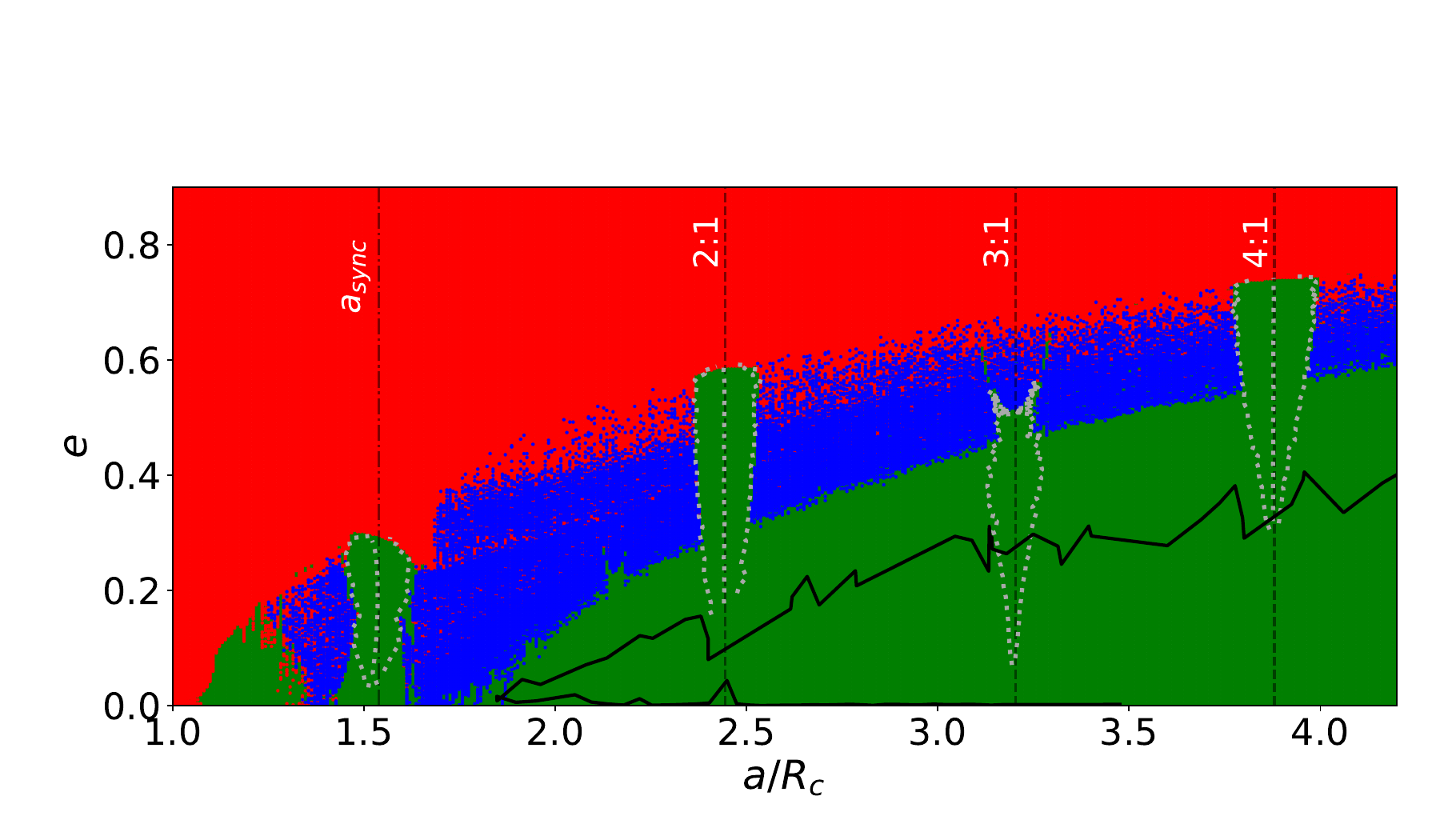}
%        \caption{$C_J = 2{,}167 \, R_c^2 \omega_k^2$.}
        \label{grade_ae_0.002_estado_COMBINADO}
    \end{subfigure}
    
    \caption{(a) Grid in the $a \times e$ plane for 160,000 particles integrated from the apocentre with $\overline{\mu} = 0.002$. Panel (a) illustrates the survival time of particles within an integration window of $10^4~T_{\mathrm{rot}}$, while panel (b) displays their final dynamical state: survivors (green), collisions (red), or ejections (blue). The white markers, represented by dotted lines in (a) and (b) denote the numerical width of the SORs (Figure \ref{gradeapore}). Furthermore, the black outline delimits the stable region associated with the boundaries of first-type orbits.}
    \label{grade_ae_0.002_estado_periodo_COMBINADO}
\end{figure}

The stability maps in the $(a,e)$ plane are shown in Figure~\ref{grade_ae_0.002_estado_periodo_COMBINADO}, where panels (a) and (b) show, respectively, the survival time and the final state of the particles. As following the PSS setup, these simulations were started with the particles at the apocentre of their orbits.

The white markers indicate the widths of the $1\!:\!1$, $2\!:\!1$, $3\!:\!1$, and $4\!:\!1$ SORs, whose boundaries were independently determined from the PSS analysis. The close correspondence between the long-lived regions in the stability maps and the resonant domains delimited by the PSS provides an independent dynamical characterisation of these regular structures and indicates that the adopted integration interval is sufficient to capture the principal dynamical structure of the system. Likewise, the boundary of the first-type orbit family closely follows the corresponding long-lived domain in the $(a,e)$ map. Trajectories within these regular domains remain stable throughout the adopted integration interval. Nevertheless, because the stability classification is necessarily finite-time, trajectories located near chaotic or sticky boundaries may remain sensitive to longer integration times.

The region inside the curve defined by the first-type orbit domain is filled by particles that survive the entire integration period. This behaviour is expected, since the technique employed delimits the regions of stability of the system based on the PSS \citep[see][]{Ribeiro2023}. However, it is important to distinguish between mere survival and dynamical stability. Although particles outside the resonance islands or the first-type orbit domain may also persist for the duration of the integration, this survival does not imply long-term stability. In reality, as previously evidenced by the Poincaré Surface of Section, these trajectories are chaotic. This chaotic nature and its long-term implications will be systematically demonstrated in the following section through the analysis of Lyapunov exponents.

%It is worth noting that the $3\!:\!1$ SOR was not superimposed on this combined analysis. For even-order SORs such as this one, the stable dynamical structures in the section plane do not align with the $\dot{x}_0=0$ condition. This implies that the orientation of these stable orbits, dictated by the argument of the pericentre, deviates from the specific configurations of $\omega_0 = 180^\circ$ and $f_0 = 180^\circ$ chosen for the grid method. Consequently, a direct fit cannot be achieved, as the uniform grid restricts the initial alignment of the particles to the main geometric axes of the coordinate system, failing to capture the rotated equilibrium points characteristic of the $3\!:\!1$ resonance.

\subsection{Chaos quantifier}
\label{SectionLyapunovResults}

We apply the LCE analysis to the initial conditions defined by the \((a,e)\) grid to assess orbital stability by quantifying the exponential divergence of nearby trajectories. While the survival maps provide the final state and survival time of each particle over the adopted integration interval, the LCE analysis provides an additional measure of the relative divergence of nearby trajectories. Following the FTLE interpretation established in the previous section, the results are discussed comparatively, focusing on differences between regions of the $(a,e)$ plane.

For this simulation, the upper limit of the colour scale is set by the largest $T_L$. The intermediate colour transitions are chosen empirically based on the statistical distribution of the $T_L$ values, allowing differences between dynamically distinct regions of the map to be represented more clearly. These colour levels are intended to improve the visualisation of the spatial variations in dynamical behaviour and do not represent fixed physical thresholds between discrete stability regimes.

In this work, we employ the LCE to characterise and quantify the chaotic nature of each simulated particle, starting from the initial conditions, early defined by the $(a, e)$ grid. This approach is primarily justified by the computational constraints associated with long-term numerical integrations. While a particle may survive the $10^4$ orbital periods defined in our grid method, this interval represents a relatively short timescale when compared to the actual lifetime required for physical structures such as rings or satellites to remain dynamically stable. Integrating the equations of motion over full secular or evolutionary timescales would be computationally unfeasible. By identifying the sensitive dependence on initial conditions within a feasible integration time, we can infer the long-term dynamical across much larger timescales than those directly simulated \citep[see][]{Madeira2026Quaoar}.

\begin{figure*}
    \centering
     \includegraphics[width=1.\linewidth]{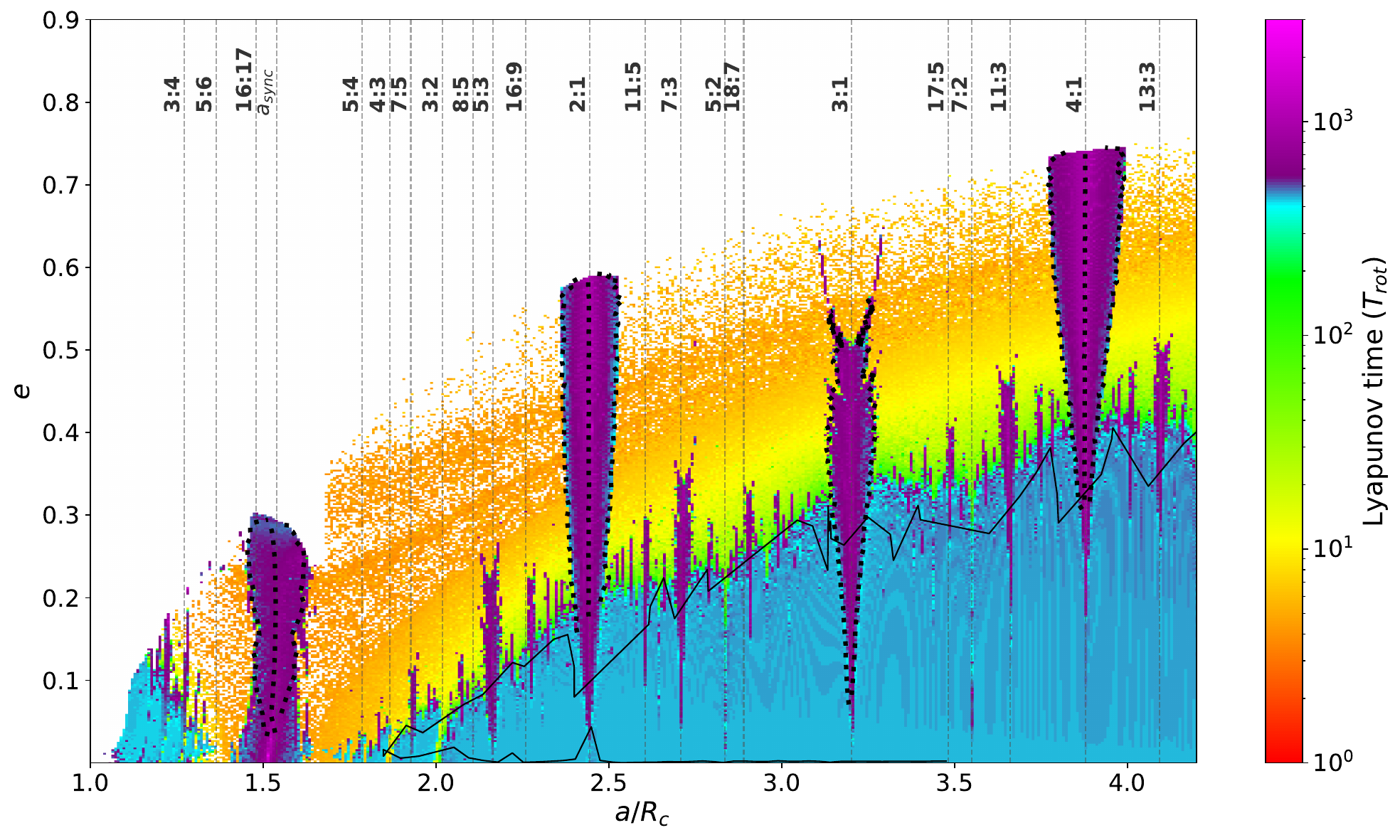}
    \caption{Stability map in the $a \times e$ plane for $\overline{\mu}=0.002$, composed of $160,000$ particles initially at their apocentre. The colour gradient indicates the time required for the loss of predictability in the particle's trajectory, measured in periods of rotation of the central body ($T_{\mathrm{rot}}$). White regions represent collided particles. The black outline delimits the stable region associated with the boundaries of first-type orbits. The black dots correspond to the width of the SORs (\ref{gradeapore}).}
    \label{fig:lyapunov}
\end{figure*}

Figure~\ref{fig:lyapunov} reveals dynamical features that are not captured by either the stability maps in Figure~\ref{grade_ae_0.002_estado_periodo_COMBINADO} or the Poincaré surfaces of section shown in Figure~\ref{fig:painel_poincare}. The white region represents particles that did not survive the full integration interval, whereas the orange and yellow regions correspond to particles with comparatively short \(T_L\). The green regions trace the chaotic layers associated with the resonant separatrices, marking the transition between domains of stronger chaotic behaviour and the more regular resonant regions. An important feature revealed by this analysis is that, although the first-type orbit domain located below the black curve and characterised predominantly by blue tones remains dynamically regular, it exhibits lower relative stability than the cores of the resonance islands highlighted in purple. This contrast may suggest that resonant confinement provides stronger protection against dynamical diffusion than the regular motion associated with the first-type orbit domain.

It is also possible to identify stability gradients within the SOR regions themselves; as expected, it can be seen that the central regions are more stable, while the edges, although still linked to SOR, exhibit libration processes that favour chaotic diffusion at slightly lower FTLE.

The $4\!:\!1$, $3\!:\!1$, and $2\!:\!1$ SORs, as well as the region near the synchronous radius, are well-defined and exhibit a high degree of correlation between the PSS and Lyapunov map techniques. These SORs stand out as the most prominent features, with the FTLE being characteristically higher within these regions. The application of the FTLE reveals dynamical details that extend beyond those obtained from PSS characterisations. Specifically, the $13\!:\!3$ and $11\!:\!3$ SORs appear stable, along with other subtle structures in the vicinity of the $4\!:\!1$ SOR. Additionally, the $7\!:\!3$ and $5\!:\!3$ SORs exhibit stable structures, while the $7\!:\!2$, $5\!:\!2$, and $3\!:\!2$ positions show small incursions of particles with lower FTLE values. In contrast, the regions corresponding to the $3\!:\!2$, $4\!:\!3$, and $5\!:\!4$ SORs are marked by distinct gaps composed of particles undergoing rapid chaotic diffusion. Further subtle stable structures are discernible at the positions of the $17\!:\!5$, $11\!:\!5$, $18\!:\!7$, $16\!:\!9$, and $7\!:\!5$ SORs, as well as the $3\!:\!4$ SOR, which is located internal to the synchronous radius. It is also noteworthy that the region near the synchronous radius may host the $1\!:\!1$ SOR, alongside various overlapping high-order commensurabilities. 

However, a definitive confirmation of the presence of particles within these specific SORs would require a systematic analysis of the respective resonant arguments to monitor their libration over time. Since such an individual orbital characterisation exceeds the scope of the present work, these structures are identified based on the dynamical correspondence between our stability profiles and the predicted resonant widths.

Note that the $3\!:\!1$ SOR, whose libration width is prominent in Figure \ref{gradeapore}, appears stable within the stability grid, and its numerical width aligns remarkably well with the results obtained via the PSS technique. In the high-eccentricity regime ($e > 0.5$), this structure exhibits a distinct dual-branch geometry that directly reflects the period-doubling bifurcation previously discussed. This behavior aligns with the general findings of \cite{celletti2026bifurcation,Celletti_2026}, who demonstrated that variations in orbital eccentricity can trigger resonance bifurcations and alter the system stability profile. However, unlike other resonance configurations where bifurcations might switch to asymmetric states outside the conventional apsides, our model captures this structural transition precisely at the apocenter ($\dot{x}_0 = 0$). Because the initial conditions of the stability grid are integrated from this exact configuration, the grid method successfully maps the full extent of this bifurcated structure, resulting in two prominent upper branches that match the underlying phase space topology of the $6\!:\!2$ SOR state.

Figure~\ref{fig:lyapunov} also reveals that the regions separating the SORs at relatively low eccentricities form a transitional chaotic layer, represented by yellow to light-green tones. These regions exhibit an intermediate dynamical behaviour between the strongly chaotic domains and the more regular regions associated with the resonant and first kind orbits structures. Notably, these particles are situated external to the curve delimiting the first-type orbit domain. It should be observed that, while the FTLE in these intermediate regions is lower than that recorded within the SOR widths, it remains distinct from the values found inside the area delimited by the first-type orbit (see Figure \ref{gradeapore}). This behaviour demonstrates that particles in SORs, particularly those close to the centre of the libration island, exhibit a significantly longer dynamic predictability time.

Monitoring the behaviour of \(\gamma\) as a function of time on a log-log scale provides a useful distinction between regular and chaotic trajectories. For regular motion, the LCE decreases approximately as \(\gamma \propto t^{-1}\), producing a slope close to \(-1\). In contrast, for chaotic trajectories, \(\gamma\) tends towards a finite positive value as the integration time increases. The corresponding FTLE, \(T_\gamma=1/\gamma\), provides an estimate of the characteristic timescale over which initially nearby trajectories diverge significantly. However, the identification of weakly chaotic trajectories may be complicated by stickiness near the boundaries of stable resonant structures, where chaotic trajectories can display regular-like behaviour for extended intervals \citep[see][]{Murray1999, ferrazmello2021caos}.

\begin{figure}
    \centering
    \includegraphics[width=1.\linewidth]{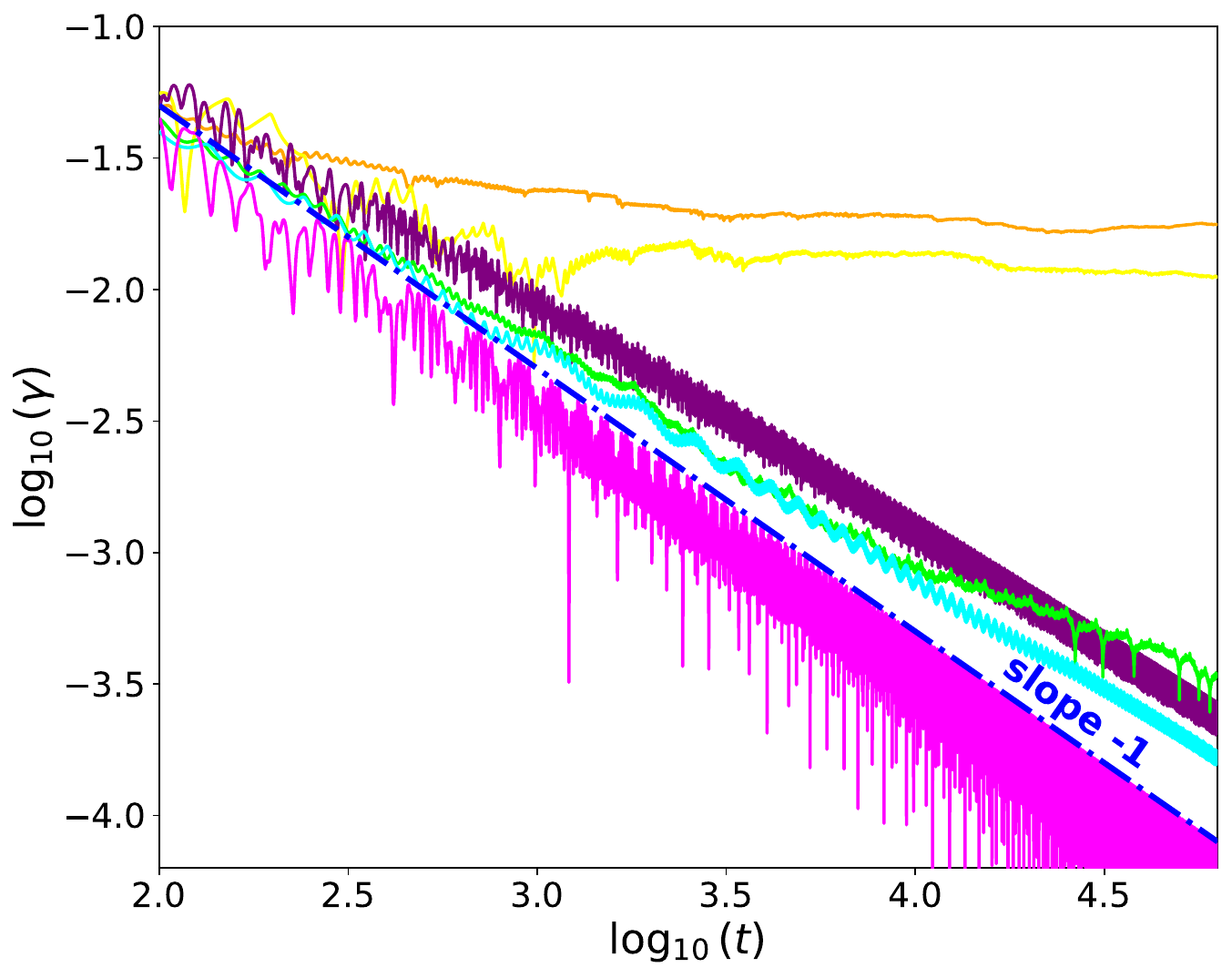}
    \caption{Temporal evolution of the finite-time largest Lyapunov characteristic exponent, \(\gamma\), on a log-log scale for representative trajectories selected from different regions of the FTLE map. The blue dash-dotted line has slope \(-1\) and represents the asymptotic \(\gamma\propto t^{-1}\) behaviour expected for regular motion. The pink \((a=1.457\,R_c,e=0.052)\), purple \((a=1.481\,R_c,e=0.172)\), cyan \((a=4.184\,R_c,e=0.381)\), and lime \((a=2.748\,R_c,e=0.237)\) trajectories display a sustained decay of the LCE, whereas the yellow \((a=3.655\,R_c,e=0.483)\) and orange \((a=1.882\,R_c,e=0.232)\) trajectories progressively approach finite positive values, indicating chaotic motion. Curves are vertically shifted for visual clarity; therefore, their absolute vertical positions should not be directly compared, while their temporal trends and slopes remain unchanged.}
    \label{fig:lyapunov_log}
\end{figure}

Figure~\ref{fig:lyapunov_log} shows the temporal evolution of the largest LCE, \(\gamma\), for representative trajectories selected from different regions of the FTLE map. The blue dash-dotted line, with slope \(-1\), represents the asymptotic \(\gamma \propto t^{-1}\) behaviour expected for regular motion. The pink \((a=1.457\,R_c, e=0.052)\), purple \((a=1.481\,R_c, e=0.172)\), and cyan \((a=4.184\,R_c, e=0.381)\) trajectories exhibit a sustained decrease of the LCE throughout the integration, broadly following the \(t^{-1}\) trend. The green trajectory \((a=2.748\,R_c, e=0.237)\) initially displays a similar decay, suggesting regular-like behaviour at shorter times; however, after \(\log_{10} t \approx 4.0\), it progressively departs from the \(t^{-1}\) trend and develops a slope characteristic of chaotic motion. In contrast, the yellow \((a=3.655\,R_c, e=0.483)\) and orange \((a=1.882\,R_c, e=0.232)\) trajectories more clearly tend towards approximately constant positive values of \(\gamma\), which is characteristic of chaotic motion. Notably, the orange trajectory corresponds to the smallest Lyapunov time in the map, with \(T_L \approx 5\,T_{\mathrm{rot}}\). The similarity between the orange and yellow curves indicates qualitatively comparable chaotic behaviour, although their asymptotic LCE values, and consequently their Lyapunov times, are different. Overall, the figure illustrates a gradual transition from trajectories whose LCE continues to decay towards the behaviour expected for regular motion to trajectories for which the LCE converges towards a finite positive value.

After establishing the characteristics of the nominal case, the following section explores scenarios involving variations in the rotation of the central body and the size of the crater. This analysis aims to determine how the dynamic structure of the system evolves in response to changes in the degree of perturbation.

\section{Varying the physical parameters} \label{Secao4}

Having characterised the nominal case based on the estimated physical parameters of Máni, we now explore how variations in the rotation rate, $\lambda$, and the crater size, $\overline{\mu}$, modify the dynamical structure in the vicinity of the system. For this purpose, FTLE maps are constructed for six different scenarios, varying one physical parameter at a time while keeping the other fixed. These simulations use a lower-resolution grid of 40,000 initial conditions per map, allowing a broader exploration of the parameter space while retaining the same chaos quantifier adopted for the nominal case.

For each simulation, the upper limit of the colour scale is set by the largest finite $T_L$ obtained among the surviving particles, while the intermediate colour transitions are selected from the empirical distribution of the corresponding $T_L$ values. Because the LCE estimator is evaluated over a finite integration interval, these scales are used to emphasise the relative dynamical organisation within each individual map rather than to define universal stability thresholds. Consequently, the colour scale is adjusted independently for each physical configuration in order to represent, in a clear and readable manner, the variations in $T_L$ among different regions of the same map. The selected colour intervals are therefore intended to highlight the relative differences in dynamical behaviour within each simulation.

For the cases in which $\lambda$ is varied, the semi-major-axis interval, $\Delta a$, is defined separately for each value so that the mapped domain includes at least the structure associated with the $4\!:\!1$ SOR. This adjustment is necessary because changes in the rotation rate produce substantial shifts in the locations of the SORs. In contrast, when $\overline{\mu}$ is varied, the corresponding shifts in resonance location are comparatively small, allowing the same $\Delta a$ interval to be retained for all mass-parameter cases.

%In this section, the analysis of the dynamics of particles around a body with a crater takes on an exploratory character, with the aim of understanding how the physical parameters $\lambda$ and $\overline{\mu}$ influence the dynamic structure of the system. The methodology employed is based on the construction of stability maps based on the maximum FTLE of each trajectory. Unlike the maps presented in Section \ref{SectionLyapunovResults}, this analysis uses a lower resolution grid, with 40,000 particles distributed on each map. For cases in which the parameter $\lambda$ is varied, the extension of the semi-major axis ($\Delta a$) on the x-axis was selected so as to cover, at a minimum, the structure corresponding to the $4\!:\!1$ SOR. Thus, $\Delta a$ is calculated specifically for each value of $\lambda$, since the rotation rate of the central body is the predominant factor in the displacement of SORs. On the other hand, for cases in which the parameter $\overline{\mu}$ is varied, although the position of the SORs undergoes corrections, these are not significant enough to require a change in $\Delta a$ in which the particles are distributed.

\begin{figure*}
    \centering

    \begin{subfigure}{\textwidth}
        \centering
        \includegraphics[width=1.\linewidth]{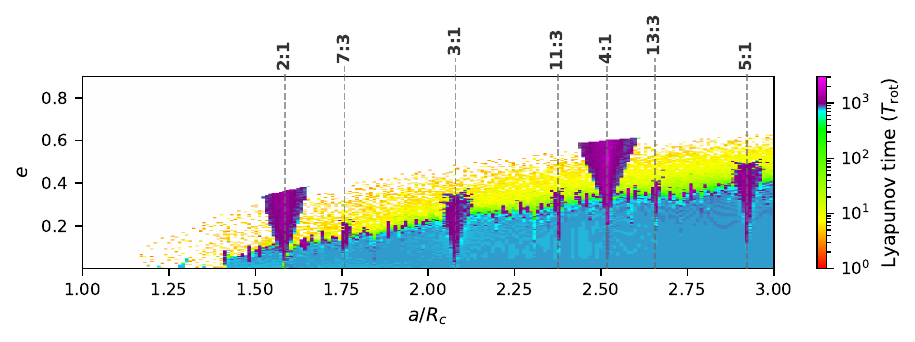}
        \caption{$\lambda = 1.0$}
        \label{fig:mu002_lbd1}
    \end{subfigure}
    \vspace{-0.12cm}

    \begin{subfigure}{\textwidth}
        \centering
        \includegraphics[width=1.\linewidth]{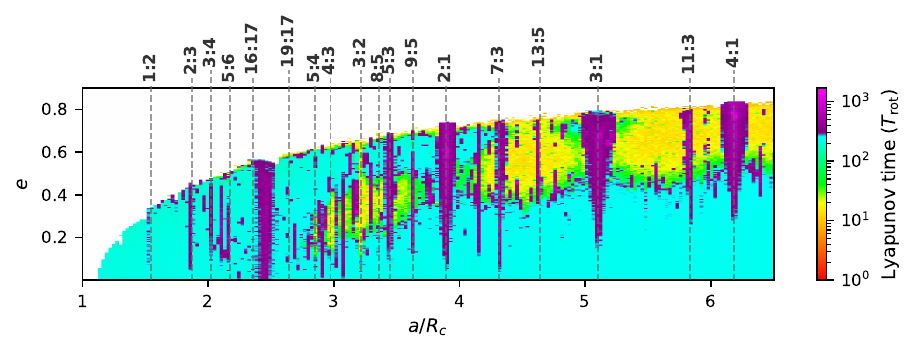}
        \caption{$\lambda = 0.26$}
        \label{fig:mu002_lbd026}
    \end{subfigure}
    \vspace{-0.12cm}

    \begin{subfigure}{\textwidth}
        \centering
        \includegraphics[width=1.0\linewidth]{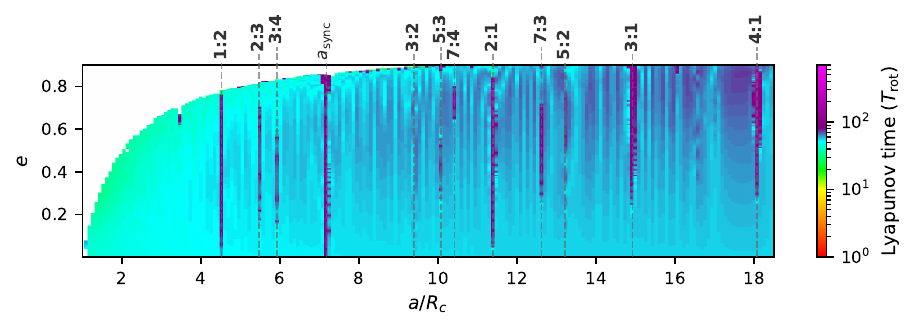}
        \caption{$\lambda = 0.052$}
        \label{fig:mu002_lbd0052}
    \end{subfigure}

    \caption{Stability maps in the $(a, e)$ plane for a fixed mass ratio $\overline{\mu} = 0.002$ and varying shape parameters: (a) $\lambda = 1.0$, (b) $\lambda = 0.26$, and (c) $\lambda = 0.052$. The color gradient indicates the time required for the loss of predictability in the particles' trajectories, measured in rotation periods of the central body ($T_{\mathrm{rot}}$). White regions represent non-surviving particles that collided with the central body or were ejected from the system during the integration period.}
    \label{fig:painel_lbd_vertical}
\end{figure*}

Figure~\ref{fig:painel_lbd_vertical} shows the stability maps based on the FTLE, considering $\overline{\mu}=0.002$ and different scenarios for the rotation period of the central body.  Figure \ref{fig:painel_lbd_vertical}~(a) details the case where $\lambda=1.0$, a value representing the theoretical rotation limit for the collapse of the central body. Under these extreme conditions, while the general stable domain is restricted to eccentricities $e \le 0.4$, the presence of resonant structures allows for stable trajectories at slightly higher eccentricities, reaching up to $e \le 0.6$. Beyond this threshold, there is a complete absence of surviving trajectories, indicating that the stability of the system under such conditions is restricted to lower eccentricities. This behaviour can be attributed to the high rotation speed of the central body, which shifts the resonances closer to the central body and enhances the overlap between different SORs. Simultaneously, the increased frequency of close encounters with the crater excites particle eccentricities, accelerating the chaotic diffusion that disrupts these trajectories. Four prominent isolated structures are identified, associated with the $2\!:\!1$, $3\!:\!1$, $4\!:\!1$ and $5\!:\!1$ SORs. In addition to these primary features, fainter structures are discernible at the theoretical positions of the $7\!:\!3$, $11\!:\!3$, and $13\!:\!3$ SORs. It is important to note that the $5\!:\!1$ SOR is visible in this specific case due to the chosen range of the semi-major axis ($\Delta a$). Consequently, its apparent absence in other scenarios with different $\lambda$ values does not necessarily imply non-existence; rather, it may indicate that the integration range for those cases was insufficient to encompass the region associated with this resonance.

When considering a reduced rotation rate ($\lambda = 0.26$), Figure~\ref{fig:painel_lbd_vertical}~(b) reveals an expansion of the region exhibiting intermediate to high dynamical regularity, extending to eccentricities of up to $e \approx 0.8$. At the same time, the distribution of chaoticity is modified by the emergence of three broad clusters characterised by comparatively stronger chaotic behaviour. These cluster-like structures, observed near $a \approx 3.3~R_c$ (for $e \in [0.1,0.45]$) and $a \geq 4.0~R_c$, are not uniformly chaotic, but contain resonant structures embedded within the surrounding chaotic domain. The regions associated with the SORs are clearly demarcated (purple and pink), with their centres occupying some of the most regular portions of the map. In this scenario, the structure associated with the synchronous radius, together with the $2\!:\!1$, $3\!:\!1$, and $4\!:\!1$ SORs, emerges as the most prominent feature. These are accompanied by less pronounced structures associated with the $7\!:\!3$, $11\!:\!3$, $5\!:\!3$, and $13\!:\!5$ SORs. Additionally, several narrow and compact structures are discernible along the semi-major axis, suggesting the presence of higher-order SORs.

In the scenario where the central body rotates more slowly ($\lambda=0.052$), as shown in Figure~\ref{fig:painel_lbd_vertical}~(c), a significant shift in the dynamical structure is observed. Notably, there is a complete absence of broad regions exhibiting rapid chaotic diffusion, as well as the disappearance of the high-chaoticity clusters that were prominent at higher rotation rates. It is noted that the population of surviving particles, extends to high eccentricities ($e \geq 0.9$). The colour gradient in the regions outside the resonant structures tends predominantly towards blue, indicating that these regions occupy the more regular portion of the dynamical range represented within this simulation. Fainter and more discrete stability lines (purple) are discernible along several commensurability lines; specifically, the $2\!:\!3$, $3\!:\!4$, $5\!:\!3$, $7\!:\!4$, $7\!:\!3$, and $5\!:\!2$ SORs continue to exhibit thin structures that indicate predictability over more restricted ranges of eccentricity. Notably, within the more prominent structures, particularly those associated with the $4\!:\!1$, $3\!:\!1$, and $2\!:\!1$ SORs, a significant number of particles reach the upper range of the Lyapunov-time scale adopted for this simulation.

It is important to note that the maximum $T_L$ ($\approx~700~T_{\mathrm{rot}}$) obtained in this case is considerably lower than the upper values reached in the other investigated configurations. This should not necessarily be interpreted as evidence of lower absolute stability. Because the conventional Lyapunov estimator is evaluated over a finite interval and converges progressively with integration time, the slowly rotating configuration may require longer integrations for the trajectories exhibiting the most regular behaviour to reach a more asymptotic estimate of their Lyapunov times. The values obtained here should therefore be interpreted in terms of the relative stability structure within this map.

The reduced rotation rate also shifts the SORs farther apart, thereby reducing the overlap between neighbouring resonances. At the same time, over the adopted integration interval, this configuration leads to fewer close encounters with the crater and consequently reduces the dynamical perturbations experienced by the particles. This more regular dynamical environment favours the preservation of a diverse set of fine resonant structures, even at high eccentricities, although the widths of the SORs decrease as the rotation rate becomes smaller. This behaviour of the resonance widths will be discussed in more detail at the end of this section.

\begin{figure*}
    \centering

    \begin{subfigure}{\textwidth}
        \centering
        \includegraphics[width=1.\linewidth]{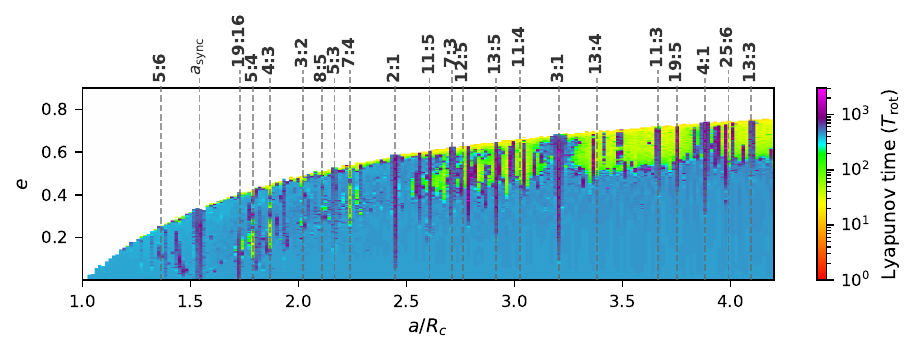}
        \caption{$\overline{\mu} = 0.0001$}
        \label{fig:lbd523_mu00001}
    \end{subfigure}
    \vspace{-0.3cm} % Espaçamento entre as figuras

    \begin{subfigure}{\textwidth}
        \centering
        \includegraphics[width=1.\linewidth]{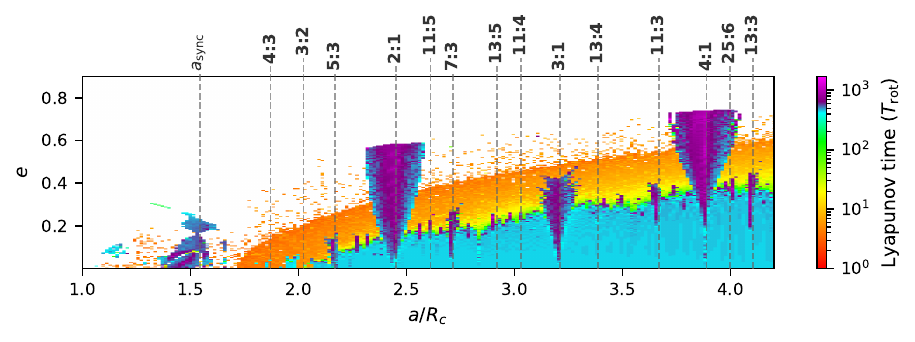}
        \caption{$\overline{\mu} = 0.005$}
        \label{fig:lbd523_mu005}
    \end{subfigure}
    \vspace{-0.3cm}
    
    \begin{subfigure}{\textwidth}
        \centering
        \includegraphics[width=1.0\linewidth]{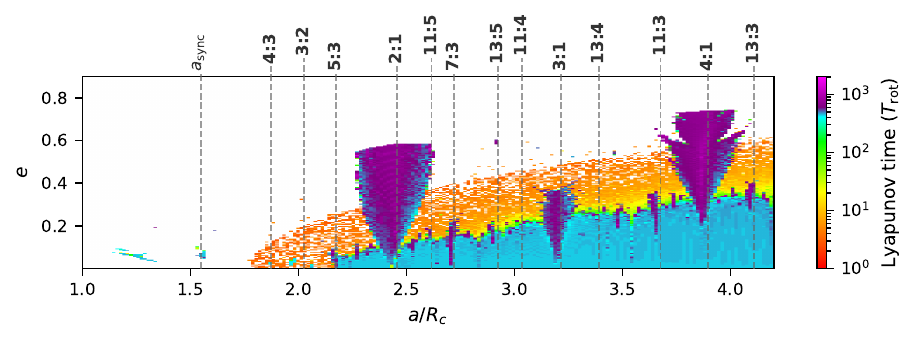}
        \caption{$\overline{\mu} = 0.01$}
        \label{fig:lbd523_mu001}
    \end{subfigure}

    \caption{Stability maps in the $(a, e)$ plane for a fixed shape parameter $\lambda = 0.523$ and varying mass ratios: (a) $\overline{\mu} = 0.0001$, (b) $\overline{\mu} = 0.005$, and (c) $\overline{\mu} = 0.01$. The color gradient indicates the time required for the loss of predictability in the particles' trajectories, measured in rotation periods of the central body ($T_{\mathrm{rot}}$). White regions represent non-surviving particles that collided with the central body or were ejected from the system during the integration period.}
    \label{fig:painel_mu_vertical}
\end{figure*}

Following the analysis of the rotation parameter $\lambda$, we now evaluate the influence of the mass ratio $\overline{\mu}$ on the dynamical structure. For this purpose, we fix the rotation rate at $\lambda = 0.523$, representing the nominal case (based on Máni parameters), and vary the mass ratio of the crater among the values $\overline{\mu} = 0.0001$, $0.005$, and $0.01$.

Figure~\ref{fig:painel_mu_vertical} presents this comparison, with panel (a) illustrating the case for $\overline{\mu} = 0.0001$, representing the smallest mass ratio considered in this work. Under this regime, numerous fine structures associated with multiple SORs are clearly identifiable by purple tones coinciding with the locations of the resonant structures. Outside these domains, the $(a,e)$ plane exhibits a comparatively more uniform intermediate level of regularity, interrupted by sparse and localised regions of enhanced chaotic diffusion.

While several resonant signatures are visible, those associated with the $1\!:\!1$, $2\!:\!1$, $3\!:\!1$, and $4\!:\!1$ SORs remain the most prominent. In this low-mass regime, these structures appear significantly thinner, consistent with the weaker perturbation produced by a smaller $\overline{\mu}$ and the consequent reduction in resonance widths. Regions exhibiting comparatively faster chaotic diffusion, represented by yellow and green tones, are found predominantly at higher eccentricities, within localised clusters, and along the commensurability lines associated with the $7\!:\!4$, $4\!:\!3$, and $5\!:\!4$ SORs. These regions indicate locally enhanced chaotic behaviour relative to the surrounding phase space, whereas the regular cores of the thin and prominent SORs mentioned above remain among the most stable structures identified in the map.

Regarding Figure~\ref{fig:painel_mu_vertical}~(b), where $\overline{\mu} = 0.005$, a significant shift in the dynamical structure is observed. In this case, the larger crater generates a more intense gravitational disturbance, leading to the emergence of gaps composed of both particles exhibiting comparatively faster chaotic diffusion (represented in orange and yellow) and particles that do not survive the full simulation period (represented in white). These gaps are notably present between a small region very close to the central body ($\approx 1.12~R_c$) and the synchronous $1\!:\!1$ SOR, as well as near $a \approx 1.7~R_c$. Additionally, smaller gaps of enhanced chaotic diffusion are found along the $4\!:\!3$ and $3\!:\!2$ resonance lines, indicating that these regions exhibit greater dynamical instability under this regime. There is also a considerable increase in particles exhibiting stronger chaotic behaviour at higher eccentricities, coupled with a decrease in the number of particles that survive the entire integration interval. The structures associated with the $2\!:\!1$, $3\!:\!1$, and $4\!:\!1$ SORs remain the most prominent, with their respective cores occupying some of the most regular regions of the map, although the $3\!:\!1$ SOR exhibits a comparatively less regular core than those associated with the $2\!:\!1$ and $4\!:\!1$ resonances. Accompanying these features, other small and fine structures remain discernible, associated with the $5\!:\!3$, $7\!:\!3$, $11\!:\!3$, $25\!:\!6$, and $13\!:\!3$ SORs. Conversely, although the region near the $1\!:\!1$ SOR contains particles exhibiting comparatively more regular behaviour than its surroundings, the resonant structure is not clearly characterised. This suggests a greater susceptibility to dynamical changes as the degree of perturbation increases, consistent with its proximity to the central body.

For the most massive case considered in this work, with $\overline{\mu} = 0.01$ (Figure~\ref{fig:painel_mu_vertical}~(c)), the structures associated with the $2\!:\!1$ and $4\!:\!1$ SORs remain predominant. Notably, these regions exhibit a visible expansion in their dynamical extent compared to the previous scenarios; such broadening is consistent with the established relationship between resonance width and the magnitude of the gravitational perturbation, as dictated by $\overline{\mu}$ and $\lambda$. Interestingly, despite this overall growth, the $3\!:\!1$ SOR does not exhibit a significant structural change, maintaining a comparatively less regular core than those associated with the $2\!:\!1$ and $4\!:\!1$ resonances. Between approximately $1.0$ and $1.75~R_c$, only a few dispersed particles survive until the end of the simulation, displaying a broad range of dynamical behaviour that includes regions of comparatively faster chaotic diffusion. Despite the increased perturbation in this scenario, several fine structures associated with the $5\!:\!3$, $7\!:\!3$, $11\!:\!3$, and $13\!:\!3$ SORs remain discernible and relatively regular, although they appear less pronounced than in the previous cases. Furthermore, regions that in the $\overline{\mu}=0.005$ case were populated predominantly by surviving particles exhibiting comparatively strong chaotic behaviour now also contain a substantial fraction of non-surviving particles. This coexistence of enhanced chaotic behaviour and particle loss indicates an increase in dynamical instability relative to the previous case, with regions that previously retained particles throughout the integration interval becoming increasingly affected by collisions or ejections.

\section{Discussion and conclusions} \label{Secao5}

%\gus{UMA COISA SUPER IMPORTANTE QUE VOCE NÃO FALOU NO TEXTO É QUE SER OU NÃO MAIS ESTAVEL VAI ALÉM DO NÚMERO QUE VOCE ENCONTRA PARA O LCE. SIGNIFICA QUE O MAIS ESTAVEL ANTE A EFEITOS EXTERIORES COMO IRREGULARIDADES NO FORMATO/PRESSÃO DE RADIAÇÃO AINDA DEVE SER O MAIS ESTÁVEL. ISSO PARECE BOBO, MAS É IMPORTANTE E TE AJUDA A CONCLUIR MUITA COISA.}

In this work, we investigated the dynamics around a spherical body with a deep crater. As a nominal case, we constructed a body with physical parameters derived from observational data of the TNO Máni ($\lambda=0.52$ and $\overline{\mu}=0.002$). Our first analysis was conducted by analysing the evolution of test particles using the Poincaré Surfaces of Section (PSSs) technique, and we then complemented the study by performing new numerical simulations with test particles distributed in a semi-major axis versus eccentricity grid. These simulations allowed us to determine the survival time of the particles, their final state, along with their Finite-Time Lyapunov Exponent (FTLE), enabling the identification of chaotic and quasi-regular regions. Additional scenarios were also explored, varying the physical parameters associated with the rotation of the central body and the size of the crater, in order to assess their influence on the dynamics of the particles in the system.

The analysis of the PSS, restricted to the nominal case, allowed the width of the $1\!:\!1$, $2\!:\!1$, $3\!:\!1$, and $4\!:\!1$ SORs to be determined \citep[see][]{winter1997resonance}. The comparison between the numerically obtained positions and those from analytical models revealed a maximum relative deviation of 2.82\% for the $1\!:\!1$ SOR, with the error progressively decreasing for resonances located further from the central body, reaching 1.34\% for the $4\!:\!1$ SOR (Table \ref{tab:ressonancias_completa}). This trend suggests that the analytical correction for the resonance position, which is based on an expansion in terms of $\overline{\mu}$ (Equations \ref{wpexpansion} and \ref{kappaexpansion}), is more affected by higher-order terms in regions closer to the primary.

The technique based on survival time proved effective in delimiting the boundaries of instability, providing a preliminary identification of regions where SORs might persist even at high eccentricities; however, as seen in applications to Centaur dynamics such as that of Chiron \citep{Madeira2025d} and also here, we note that this metric is insufficient to fully evaluate long-term dynamic resilience. This limitation arises because chaotic trajectories with slow diffusion rates may persist for the entire simulation period without experiencing macroscopic instabilities, thereby masking the underlying chaotic nature of the region. Conversely, analysis via FTLE successfully reveals the internal nuances of these apparent stability zones, allowing the widths of the SORs to be precisely measured and distinct levels of regular and irregular motion to be differentiated through explicit stability gradients. This approach clearly highlighted the chaos in the SOR separators, a consequence of the SOR overlap mechanisms. Notably, the degree of stability identified by the FTLE goes beyond the numerical value of the LCE; a region characterised as more stable implies a higher level of dynamical robustness against external perturbations, such as non-modelled spherical harmonics ($J_2$ and $C_{22}$) or non-conservative mechanisms, including solar radiation pressure and the viscous dissipative effects already modelled by \citet{madeira2025formation}. Consequently, these low-chaos zones represent the most likely environments where material could be preserved over long timescales despite unmodelled physical complexities.

It is also important to emphasise how the $3\!:\!1$ SOR, and even-order resonances in general, are characterised by grid-based methods such as survival time and FTLE. The grid simulations are constructed under a specific set of initial orbital elements ($I = 0^\circ$, $\varpi_0 = 180^\circ$, $f_0 = 180^\circ$, and $\Omega_0 = 0^\circ$). For the PSS to strictly follow and reflect this identical set of orbital elements, all considered SOR structures were evaluated precisely at $\dot{x}_0 = 0.0$. While the PSS can characterise SOR structures for each value of $C_J$ across the entire phase space without being restricted to $\dot{x}_0 = 0$, each individual stability grid is fundamentally restricted to probing the evolution of a single, unique set of initial orbital elements. Under these conditions, the grid maps reveal a distinct bifurcation pattern at the top of their respective structures at the highest eccentricities. This split-path geometry represents a direct manifestation of the period-doubling bifurcation ($3\!:\!1 \rightarrow 6\!:\!2$ SOR) taking place exactly at this apocentre configuration (that is, $\dot{x}_0 = 0$). Consequently, for the specific set of initial conditions adopted in this work, the upper boundaries of even-order SORs naturally exhibit these bifurcated features at elevated eccentricities, a behaviour that underscores how the visible architecture of these resonances depends heavily on the initial orbital phase and orbital orientation.

Although PSS also indicate the presence of chaotic regions in the separators, the precise delimitation of the boundaries and widths of the structures associated with SORs requires exhaustive manual analysis, especially since it requires the inspection of multiple graphs generated for different Jacobi's Constant. Given the coexistence of several islands associated with SORs that evolve according to the $C_J$ of the system, the FTLE-based technique proved to be particularly efficient in automatically characterising the morphology of the dynamics and variations in these widths without the need for repeated manual processes for different $C_J$ values.

The numerical results are consistent with the analytical prediction that the parameters $\overline{\mu}$ and $\lambda$ are directly proportional to the SOR widths, as established by \cite{Madeira2022}. However, the dynamical consequences of varying these parameters differ significantly. As the rotation rate increases towards the critical velocity ($\lambda \to 1$), the maximum eccentricity of surviving particles decreases significantly, restricting stability to lower limits. Conversely, a tenfold reduction in rotation ($\lambda=0.052$) shifts the SORs further apart, minimising resonance overlap and allowing for survival at considerably higher eccentricities ($e \geq 0.9$). This trend regarding the maximum eccentricity is not replicated when varying the mass ratio; surviving SORs in high-$\overline{\mu}$ scenarios still harbour particles with high eccentricities. Instead, increasing $\overline{\mu}$ intensifies the gravitational disturbance and progressively expands the unstable regions where particles do not survive the entire integration period, while simultaneously emptying the regions closest to the central body. As the degree of perturbation rises, the maps reveal fewer but larger resonant structures, which serves as a clear numerical indicator of resonance overlap. Under these severe mass conditions, only the most prominent structures tend to persist. The $2\!:\!1$ and $4\!:\!1$ resonances, in particular, exhibit a visible broadening as $\overline{\mu}$ increases, whereas the $3\!:\!1$ SOR does not follow this expansion, a result that underscores its high sensitivity to the initial orbital orientation of the particles. It is important to emphasise that, in this work, we adopted the proportions of the largest crater ever observed as our physical baseline, based on the estimated data from \cite{Rommel_2023}; according to our estimates, this corresponds to $\overline{\mu} = 0.002$. While a crater of greater magnitude (such as $\overline{\mu} = 0.01$) may not be physically feasible in a real system, this extreme scenario was included as part of an exploratory analysis to investigate the asymptotic limits of the dynamical perturbations and to better understand the stability transitions within the system.

Regarding the implications for the Máni system, our PSS analyses indicate that the $1\!:\!1$, $2\!:\!1$, $3\!:\!1$, and $4\!:\!1$ SORs are the most prominent resonant structures. Given the recent discoveries of ring systems near the $3\!:\!1$ resonance in other TNOs and Centaurs, Máni emerges as a strong candidate for hosting rings or satellites within these stable regions, particularly around the $2\!:\!1$, $3\!:\!1$, and $4\!:\!1$ SORs.
%, since their positions relative to the Roche limit of the system support the structural longevity of such features.

Furthermore, our model represents the crater as a point of removed mass that shifts the centre of mass relative to the body's spin axis. While this approach serves as a simplified representation of a crater, it more broadly explores the dynamical consequences of a non-coincident spin axis and centre of mass, regardless of whether such an offset arises from topographical depressions or internal density variations. A key distinction arises when comparing our results with mass anomaly models motivated by ringed TNOs \citep{sicardy2019ring, celletti2026dynamical}. While those models suggest that torques from non-spherical bodies can sweep the surrounding disc and render the $2\!:\!1$ SOR unstable on short timescales, our simulations reveal the existence of stable particles in this inner region over $10^4$ rotations. This indicates that the dynamical environment produced by this mass distribution specifically allows for the persistence of stable zones where a mass anomaly would typically deplete them.

%\gus{* Tem alguma outra implicação dos seus resultados?}

\section*{Acknowledgements}

PVSS thanks CAPES (Financial Code 001) and FAPESP (Proc.~2026/08614-8). SMGW thanks FAPESP (Proc.~2022/11783-5), CNPq (Proc.~309057/2025-6) and Capes (Financial Code~001) for the financial support. OCW thanks FAPESP (proc.~2022/11783-5), CNPq (proc.~316991/2023-6). Computational resources were provided by the Center for Scientific Computing (NCC/GridUNESP) of the São Paulo State University (UNESP) and the Center for Mathematical Sciences Applied to Industry (CeMEAI), funded by FAPESP (grant 2013/07375-0). In preparing this work, the authors utilised Gemini Pro to enhance the readability of the English. Following this, the authors carefully reviewed and edited the content, assuming full responsibility for the publication’s material.

%%%%%%%%%%%%%%%%%%%%%%%%%%%%%%%%%%%%%%%%%%%%%%%%%%
\section*{Data Availability}

The data underlying this article will be shared on reasonable request to the corresponding author.
 
%The inclusion of a Data Availability Statement is a requirement for articles published in MNRAS. Data Availability Statements provide a standardised format for readers to understand the availability of data underlying the research results described in the article. The statement may refer to original data generated in the course of the study or to third-party data analysed in the article. The statement should describe and provide means of access, where possible, by linking to the data or providing the required accession numbers for the relevant databases or DOIs.



\bibliographystyle{mnras}
\bibliography{example} % if your bibtex file is called example.bib

%\begin{thebibliography}{99}
%\end{thebibliography}

%%%%%%%%%%%%%%%%%%%%%%%%%%%%%%%%%%%%%%%%%%%%%%%%%%

%%%%%%%%%%%%%%%%% APPENDICES %%%%%%%%%%%%%%%%%%%%%

%\appendix

%\section{Duplicação de período}

%If you want to present additional material which would interrupt the flow of the main paper, it can be placed in an Appendix which appears after the list of references.

%\pv{Será uma histerese que surge do fato de estar próximo ao limite caótico ?  .......... CJ = 1.800 ress $2\!:\!1$ ---- OSCILADOR DE DUFFING (estrutura muito semelhante) --- as orbitas saem e voltam para o ponto de equilíbrio instável}
%Será uma histerese que surge do fato de estar próximo ao limite caótico ? Eu gostaria de explorar mais esse topico em um trabalho separado, o que acham? 

%%%%%%%%%%%%%%%%%%%%%%%%%%%%%%%%%%%%%%%%%%%%%%%%%%

% Don't change these lines
\bsp	
\label{lastpage}
\end{document}